\documentclass[10pt,journal,comsoc]{IEEEtran} 
\usepackage{multirow}
\usepackage[textwidth=1cm,backgroundcolor=yellow,textsize=tiny]{todonotes}

 \usepackage{amsmath,amsfonts, amsthm, bm}

\usepackage{comment, tagging}

\usetag{long} 

\usepackage{colortbl}
\usepackage{stackengine} 

\usepackage{array}
\newcolumntype{P}[1]{>{\centering\arraybackslash}p{#1}}
\newcolumntype{M}[1]{>{\centering\arraybackslash}m{#1}}
\usepackage{setspace}

\usepackage[compress]{cite}
\usepackage{tabto}
\usepackage[noend]{algpseudocode}
\usepackage{algorithm}
\usepackage{varwidth} 
\usepackage{xspace, paralist}
\usepackage{url}
\PassOptionsToPackage{hyphens}{url}
\usepackage{float}
\usepackage{graphicx, pstricks, pst-node,pst-plot}
\usepackage{subfloat, tabu, color, subfiles}
\usepackage[caption=false,labelfont=sf,textfont=sf]{subfig}
\newcommand{\subfloatWidth}{0.66\columnwidth}
\newcommand{\wideSubfloatWidth}{0.98\columnwidth}
\usepackage{textcomp, stfloats, verbatim}

\algnewcommand{\AND}{\textbf{and}\xspace}
\algnewcommand{\OR}{\textbf{or}\xspace}
\algnewcommand\algorithmicforeach{\textbf{for each}}
\algdef{S}[FOR]{ForEach}[1]{\algorithmicforeach\ #1\ \algorithmicdo}

\algrenewcommand\algorithmicindent{1.0em}%

\makeatletter
\algnewcommand{\LineComment}[1]{\Statex \hskip\ALG@thistlm \(\triangleright\) #1}
\makeatother

\newbox\statebox
\newcommand{\myState}[1]{%
    \setbox\statebox=\vbox{#1}%
    \edef\thealgruleheight{\dimexpr \the\ht\statebox+1pt\relax}%
    \edef\thealgruledepth{\dimexpr \the\dp\statebox+1pt\relax}%
    \ifdim\thealgruleheight<.75\baselineskip
        \def\thealgruleheight{\dimexpr .75\baselineskip+1pt\relax}%
    \fi
    \ifdim\thealgruledepth<.25\baselineskip
        \def\thealgruledepth{\dimexpr .25\baselineskip+1pt\relax}%
    \fi
    \State #1%
    \def\thealgruleheight{\dimexpr .75\baselineskip+1pt\relax}%
    \def\thealgruledepth{\dimexpr .25\baselineskip+1pt\relax}%
}

\newcommand{\abs}[1]{\left\vert#1\right\vert}
\newcommand{\set}[1]{\left\{#1\right\}}

\usepackage{mathtools, nccmath}
\DeclarePairedDelimiter\ceil{\lceil}{\rceil}

\DeclarePairedDelimiter\floor{\lfloor}{\rfloor}
\usepackage{stackengine}

\newcommand{\setCaches}{\mathcal{N}} 
\newcommand{\numCaches}{N} 
\newcommand{\numPosInds}{n_x} 
\newcommand{\mS} {S}   
\newcommand{\cacheSize}{C}
\newcommand{\subsetCaches}{D}
\newcommand{\ind}{\tilde{I}} 
\newcommand{\indSize}{I} 

\newcommand{\indMinSize}{\indSize_{\min}}
\newcommand{\indMaxSize}{\indSize_{\max}}
 
\newcommand{\uInterval}{U} 
\newcommand{\uIntervalMinFeasible}{U_{min}} 
\newcommand{\bwBudget}{B}
\newcommand{\accsCost} {c}   
\newcommand{\period} {R}   

\newcommand{\mrone}{\pi} 
\newcommand{\mrzero}{\nu} 
\DeclareMathOperator{\init}{init}
\newcommand{\mrzeroInit}{\mrzero_{\init}} 
\newcommand{\mronej}{\mrone_j} 
\newcommand{\mr}{\rho} 
\newcommand{\mrj}{\mr_j} 
\newcommand{\mrzeroj}{\mrzero_j} 
\newcommand{\missp}{M} 
\newcommand{\costhetero}{\phi} 

\newcommand{\salsa}{SALSA2} 
\newcommand{\pif}{Opt} 
\DeclareMathOperator{\fna}{FNA}

\newcommand{\pgmfna}{HeCS$_{\fna}$} 

\date{}

\author{Itamar Cohen~\IEEEmembership{Member,~IEEE}
\thanks{
\noindent I. Cohen is with 
The Department of Computer Science, Ariel University, Ariel 40700, Israel; e-mail: itamarc@ariel.ac.il.

\noindent An early version of this work was published in~\cite{SALSA2_poster}.

\noindent This work has been submitted to the IEEE for possible publication. Copyright may be transferred without notice, after which this version may no longer be accessible.
} 
}
\begin{document}

\title{Bandwidth Efficient Cache Selection and Content Advertisement}
\maketitle

\begin{abstract}
Caching is extensively used in various networking environments to optimize performance by reducing latency, bandwidth, and energy consumption. To optimize performance, caches often advertise their content using indicators, which are data structures that trade space efficiency for accuracy. However, this tradeoff introduces the risk of false indications. Existing solutions for cache content advertisement and cache selection often lead to inefficiencies, failing to adapt to dynamic network conditions. This paper introduces SALSA2, a Scalable Adaptive and Learning-based Selection and Advertisement Algorithm, which addresses these limitations through a dynamic and adaptive approach. SALSA2 accurately estimates mis-indication probabilities by considering inter-cache dependencies and dynamically adjusts the size and frequency of indicator advertisements to minimize transmission overhead while maintaining high accuracy.
Our extensive simulation study, conducted using a variety of real-world cache traces, demonstrates that SALSA2 achieves up to 84\% bandwidth savings compared to the state-of-the-art solution and close-to-optimal service cost in most scenarios. These results highlight SALSA2's effectiveness in enhancing cache management, making it a robust and versatile solution for modern networking challenges.

\end{abstract}

\section{Introduction}

\IEEEPARstart{C}aching are extensively used in multiple networking domains including wide-area networks~\cite{summary_cache}, content delivery networks~\cite{CDN_theory_Vs_practice,Joint_opt}, 
in-network caching~\cite{ In_netw_caching_security_tnsm21,CCN_tnsm_21}, and wireless networks~\cite{uIntervalInMANET, Digest_in_Manet}.
In such networks, accessing caches incurs some overhead in terms of latency, bandwidth, or energy~\cite{Joint_opt, BloomParadox}. On the other hand, failing to retrieve a datum from a cache incurs a larger {\em miss penalty}, e.g., for fetching the requested item from a remote server~\cite{CDN_theory_Vs_practice}. 

To optimize performance, caches often advertise their content~\cite{summary_cache, CDN_theory_Vs_practice,Joint_opt, indicators_in_NDN19,ICN_survey_15,
Digest_in_Manet}.
Such advertisements allow clients to minimize costs by selecting which cache to access for a requested datum.
Due to space, energy, and bandwidth constraints, systems often compromise some accuracy for efficiency by advertising periodical approximate {\em indicators}. 
Indicators are data structures that trade space efficiency for accuracy (e.g., Bloom filters~\cite{Bloom, CBF, Survey18, BloomParadox}).
The compromised accuracy introduces a risk of false-indications, 
which result in 
unnecessary 
misses.

Traditional approaches to cache content advertisement involve fixed static strategies that do not adapt to the dynamic nature of network traffic. Such static methods can result in either excessive bandwidth consumption or suboptimal cache utilization, affecting the system's overall performance. Moreover, existing solutions discard inter-cache dependencies, where the presence of data in one cache affects the likelihood of its presence in another. Ignoring these dependencies leads to inaccurate estimation of the false-indication probability, further reducing system efficiency.

\paragraph*{Our contributions} 
we start with a taxonomy of state-of-the-art cache selection and content-advertisement algorithms, thus providing insights into the pitfalls of existing solutions. We then use these insights to develop \salsa\ - a Scalable Adaptive and Learning-based Selection and Advertisement Algorithm, which is the first holistic solution for the cache selection and cache content advertisement problems. 
\salsa\ utilizes a lightweight learning method to efficiently estimate the false indication probability while considering inter-cache dependencies to improve accuracy. Our simulations, based on real-world cache traces, show that this technique provides estimations very close to the measurements produced by an oracle that always knows whether a given indication is correct.
Then, \salsa\ integrates these false indication probabilities into known polynomial-time cache selection algorithms, to reduce costs.
Furthermore, \salsa\ dynamically scales the size and frequency of indicator advertisements based on the current workload, thus obtaining high advertisement accuracy while using low transmission overhead. In particular, \salsa\ considers advertising either the full indicator or small frequent {\em delta updates}, which list only the changes since the last advertisement, according to the perceived workload and the reliability of the transmission channel. We evaluate \salsa's performance in various system settings through extensive evaluation study using real-world traces. The results show significant reductions in service cost and bandwidth consumption, highlighting the effectiveness of our approach in various network scenarios.

The rest of the paper is organized as follows. After introducing the system model in Section~\ref{sec:model}, we discuss related work in Sec.~\ref{sec:related_work}. In Section~\ref{sec:epe}, we show how \salsa\ estimates the mis-indication probabilities. Sec.~\ref{sec:cs} details \salsa's cache selection scheme. In Sec.~\ref{sec:advertise_alg}, we detail \salsa's content-advertisement algorithm. Sec.~\ref{sec:sim} assesses the performance of \salsa\ compared to state-of-the-art solutions. Finally, Sec.~\ref{sec:conclusion} draws some conclusions.

\section{System Model}\label{sec:model}

This section formally defines our system model and notations, which are summarized in Tab.~\ref{tab:notation}.

\noindent We consider a set $\setCaches$ of $\numCaches = |\setCaches|$ caches, containing possibly overlapping sets of
items.
Let $\mS_j$ denote the set of items stored at cache $j$ (at some point in time). 
Each cache $j$ maintains an {\em indicator} $\ind_j$, which approximates the set of items in cache $j$.
$\ind_j (x) = 1$ is referred to as a {\em positive indication} while $\ind_j (x) = 0$ is considered a {\em negative indication}. The number of positive indications for requested datum $x$ is denoted $\numPosInds$.

In what follows, when estimating the probabilities affecting the performance of our system, we consider these probabilities with respect to an arbitrary item $x$ being drawn from the distribution of items in the request sequence.

For every cache $j$, let $\mrone_j = \Pr (x \notin \mS_j | \ind_j (x) = 1)$ denote the {\em positive exclusion probability}, namely, the probability that a requested arbitrary item $x$ is not in the cache, despite a positive indication.
Similarly, let $\mrzero_j = \Pr (x \notin \mS_j | \ind_j (x) = 0)$ denote the {\em negative exclusion probability}, namely, the probability that a requested arbitrary item $x$ is not in the cache, given a negative indication.

Accessing cache $j$ incurs some predefined {\em access cost}, $\accsCost_j$. 
The overall access costs of accessing a set $D$ of caches is $\accsCost_{D} = \sum_{j \in D} \accsCost_j$. 

A multi-cache data access is considered a {\em hit} if the item $x$ is found in at least one accessed cache, and a {\em miss} otherwise. 
A miss incurs a {\em miss penalty} of $\missp$, for some $\missp > \max_j c_j$.
The {\em miss cost} of access to a set of caches $\subsetCaches$ captures the expected cost of a miss, namely, the miss penalty, times the probability of a miss. 
Assuming that the exclusion probabilities of distinct caches are mutually independent\footnote{We will later revisit this assumption.}, the miss cost for a query $x$ is
$\missp \cdot \prod_{\substack{j \in \subsetCaches\\ \ind_j(x) = 1}} \mrone {j} \cdot \prod_{\substack{j \in \subsetCaches\\ \ind_j(x) = 0}} \mrzero {j}$.
The (expected) {\em service cost} of a query is the sum of the access cost and the miss cost, namely, 
\begin{equation}\label{eq:def_service_cost}
\begin{split}
    \costhetero_x (\subsetCaches)  = 
    \sum\nolimits_{j \in \subsetCaches} \accsCost_j + \missp
    \prod_{\substack{j \in \subsetCaches\\ \ind_j(x) = 1}} \mronej
    \cdot \prod_{\substack{j \in \subsetCaches\\ \ind_j(x) = 0}} \mrzeroj.
    \end{split}
\end{equation}

We let $\indSize_j$ denote the size of indicator $\ind_j$ in bits.
To use only feasible sizes, the indicator size should be within some predefined range $[\indMinSize, \indMaxSize]$.

The {\em update interval} $\uInterval$ is the number of insertions of new items that the cache counts before it advertises a fresh indicator.
The update interval is at least $\uIntervalMinFeasible$.
One could use $\uIntervalMinFeasible = 1$. However, a slightly higher interval allows piggybacking indicator updates on packets carrying cached data payloads to decrease the transmission overheads~\cite{Digest}. Each advertisement can include either a full indicator, or a {\em delta update} that lists the changes since the last advertisement. An {\em advertisement algorithm} determines the update interval $\uInterval$, the indicator size $\indSize$, and the type of advertisement (full indicator or delta update). The {\em communication overhead} is the overall number of bits transmitted for content-advertisement along the trace.

We will be considering three closely related sub-problems:
\begin{itemize}
    \item The {\em exclusion probabilities estimation problem}: estimate the exclusion probabilities $\mrone_j$ and $\mrzero_j$ for each $1 \leq j \leq \numCaches$.
    \item The {\em Cache selection problem}: Assuming that $\vec{\mrone}$ and $\vec{\mrzero}$ are known, select a set of caches that minimizes~\eqref{eq:def_service_cost}.
    \item The {\em Cache-content advertisement problem}: minimize the communication overhead.
\end{itemize}
To simplify expressions throughout the paper, we always use logarithms of base 2, and omit the base of the logarithm.
Furthermore, we typically refer to a single client and a single cache. Correspondingly, we omit the subscript $j$ when clear from the context. However, all the clients run the same cache-selection algorithm, and all the caches run the same indicator advertisement algorithm.

\begin{table}[t]
	\caption{\label{tab:notation}List of Symbols.}
    \footnotesize	 	\begin{tabular}{|l|l|}   
		\hline
		Symbol & Meaning \tabularnewline
		\hline
		$\setCaches$ & Set of caches\tabularnewline
		\hline
		$\numCaches$ & Number of caches: $\numCaches = \abs{\setCaches}$ \tabularnewline
		\hline
		$\numPosInds$ & Number of positive indications for requested datum $x$ \tabularnewline 
		\hline
		$\mS_j$ & The set of data items in cache $j$\normalsize \tabularnewline
		\hline
        $\indSize_j$ & Size of indicator $j$ \tabularnewline
		\hline
		$\indMinSize$ & Min feasible indicator size \tabularnewline
		\hline
		  $\indMaxSize$ & Max feasible indicator size \tabularnewline
		\hline
            \rule{0pt}{8pt}
		$\ind_j (x)$ & Indication of indicator $\ind_j$ for item $x$ \tabularnewline
		\hline
        $\uInterval$ & Update interval [number of inerstions]\tabularnewline 
		\hline
		  $\uInterval_{min}$ & Minimal update interval [number of inerstions] \tabularnewline
		\hline
        $\mronej$
        & Probability of a miss in cache $j$ given a positive indication\tabularnewline
		\hline
        $\mrzeroj$ 
        & Probability of a miss in cache $j$ given a negative indication \tabularnewline
		\hline
        $\accsCost_j$ & Access cost of cache $j$ \tabularnewline
		\hline
		$\missp$  & Miss penalty \tabularnewline
		\hline
		\rule{0pt}{2ex}    
		$\costhetero$  & Cost function~\eqref{eq:def_service_cost}.
        \tabularnewline
		\hline
		$\delta_{\mrzero}$ & Smoothness parameter of $\mrzero$'s moving average~\eqref{eq:estimate_mrzero}\tabularnewline
		\hline
		$\delta_{\mrone}$ & Smoothness parameter of $\mrone$'s moving average~\eqref{eq:estimate_mrone}\tabularnewline
		\hline
        $\mrj$ & Probability of a miss in cache $j$ given its indication \tabularnewline
        \hline
        $\mathcal{C}$ & Clamp on the update interval  (Alg.~\ref{alg:after_insertion:full} ln.~\ref{alg:after_insertion:full:clamp_if})\tabularnewline
		\hline 
        \end{tabular}
        \normalsize
\end{table}

\section{Related Work}\label{sec:related_work}

\subsection{Cache-Indicators Systems}

Indicators are used to periodically advertise the cache content in multiple networking environments, including wide-area networks~\cite{summary_cache}, content delivery networks~\cite{CDN_theory_Vs_practice,Joint_opt}, 
in-network caching~\cite{ In_netw_caching_security_tnsm21,CCN_tnsm_21}, and wireless networks~\cite{uIntervalInMANET, Digest_in_Manet}.
Indicators use randomized hash-based data structures such as Bloom filters~\cite{Bloom, CBF}, and fingerprint hash tables~\cite{TinyTableJournal, TinySet}. 

Advertising indicators exhibits an inherent tradeoff between accuracy and transmission overhead. Several studies focused on reducing the transmission overhead of indicators by using sophisticated data structures. For instance, transmitting compressed indicators, at the expense of larger local memory consumption and computational work in the transmitting node~\cite{CompressedBF}. 
Another approach is to accurately advertise important information while allowing less critical data to be stale, or less accurate~\cite{FPfree_Ori, HBA_journal}.  
The work~\cite{Survey18} surveys many optimizations to indicators, such as the support for removals and dynamic scaling. 

\subsection{Static Cache Selection and Advertisement Policies}

Traditionally, the impact of the cache selection scheme was largely overlooked. For instance, Summary Cache~\cite{summary_cache} suggested selecting ``caches with positive indications'' without details on which caches with positive indications to pick. Other early works~\cite{Survey12, Digest} suggested selecting the cheapest cache with a positive indication.

Production distributed caching systems often use naive and static schemes. For instance, by default, Squid cache~\cite{SquidSpec} advertises an indicator once in a fixed one-hour interval. 
Squid's spec notes that ``more data and experience is required before other periods, whether fixed or dynamically varying, can `intelligently' be chosen.''~\cite{SquidSpec}
Apache Phoenix~\cite{ApacheCache, ApacheCacheParams} allows the user to tune the parameters that determine the indicator's size and the update interval. 
Some works~\cite{summary_cache, 
uIntervalInMANET} 
focused on optimizing the advertisement policy for some concrete settings (workload, cache size, cache policy, miss penalty). Other works~\cite{b1,b5} addressed the problem of maintaining a bandwidth budget when sending advertisements using a trial-and-error approach.
\subsection{Dynamic Cache Selection and Advertisement Algorithms}\label{sec:related_work:taxonomy}

Several studies showed that the traditional static approach results in far-from-optimal performance, thus persuading the evolution of a more systematic and algorithmic approach. The Bloom Paradox~\cite{BloomParadox} coins a scenario when relying naively on indicators for the cache selection may degrade the performance, thus opening the door to more educated cache selection algorithms. The work~\cite{CAB} showed that there is no ``one-fits-all'' static advertisement algorithm. 

\begin{table*}
\caption{\label{tab:taxonomy}Taxonomy of existing solution for the three related sub-problems discussed in this paper: (I) Exclusion probability estimation, (ii) Cache selection, and (iii) Cache-content Advertisement}
\small
\begin{tabular}{|l|l||l|l|l||l||P{1cm}|P{1cm}|}
\hline
\multirow{2}{*}{Solution} & \multirow{2}{*}{\# Caches} & \multicolumn{3}{c||}{Exclusion Probability Estimation} & Cache Selection & \multicolumn{2}{P{2cm}|}{Cache-content Advertisement} \\ \cline{3-5}  \cline{7-8}
                          &                            & \multicolumn{1}{P{1.5cm}|}{FN awareness}   & \multicolumn{1}{l|}{Approach} &    \multicolumn{1}{P{3cm}||}{Inter-cache dependency awareness} &                 & uInterVal scaling  & Indicator scaling \\ \hline \hline
    \cite{CAB}       & 1     & No & No & No & No 		& Yes 		& Yes 	\tabularnewline \hline  \cite{BloomParadox}    & 1     & No 	& Popularity           & No 	& Binary decision 		& No 		& No  	\tabularnewline \hline
    \cite{Whether}      & 1     &  Yes	& Popularity           & No 	& Binary decision  	& Yes 		& No 	\tabularnewline \hline
    \cite{fpr_fnr_in_dist_replicas}   & Multiple  &  Yes 	& Indicator analysis   &No	&         No        	& No 		& Yes 	\tabularnewline \hline
     \cite{Accs_Strategies_ToN}       & Multiple  &  No  	& History              &No   	& Approx. algorithm 					& No 		& No  	\tabularnewline \hline
    \cite{chen2020sequential}       & Multiple  & No & No  & No  	& Approx. algorithm 	& No   		& No 	\tabularnewline \hline
    \cite{FN_aware_ToN}    & Multiple  &  Yes 	& Indicator analysis   & No   	& Approx. algorithm 	& No 		& No  	\tabularnewline \hline
    \salsa   & Multiple  &  Yes 	& History + learning              & Yes 	& Approx. algorithm 	& Yes   	& Yes 	\tabularnewline \hline
\end{tabular}
\end{table*}

Tab.~\ref{tab:taxonomy} presents a taxonomy of state-of-the-art solutions for the three sub-problems we study, namely: (i) the exclusion probability estimation problem, (ii) the cache selection problem, and (iii) the cache-content advertisement problem. 
When a paper disregards a problem, the relevant entry says merely "No." 
As shown in the table, some of the papers consider a single-cache system. While providing some important insights, these solutions cannot be easily generalized to the multi-caches environment. For instance, in the particular case of a single cache, the cache selection problem is a mere binary decision -- whether to access the cache or not -- while in a multi-caches environment, it is an NP-hard problem~\cite{chen2020sequential}. 
We now survey the existing solutions for each of the sub-problems.

\paragraph*{Exclusion Probability Estimation}
Some studies~\cite{Accs_Strategies_ToN} overlook false negative indications. However,~\cite{FN_aware_ToN, CAB, Whether, fpr_fnr_in_dist_replicas} showed that practical systems exhibit significant false-negative probability, which may translate to a crucial performance degradation. Several approaches exist to estimate the exclusion probability. 
The studies~\cite{BloomParadox, Whether} calculated the exclusion probability of each queried item using its (estimated) popularity. However, accurately estimating items' popularity is challenging even for the single-cache framework used by~\cite{BloomParadox, Whether}, and may be impossible for a multiple-cache system. 
Another approach relies on a combinatorial analysis of the number of bits in the counter that have been set/reset since the last advertisement~\cite{fpr_fnr_in_dist_replicas, FN_aware_ToN}. 
However, this approach implicitly assumes that the workload is drawn from a uniform distribution. 
Our experimental study in Sec.~\ref{sec:epe:sim} shows that this assumption translates to significant estimation errors. 
~\cite{FN_aware_ToN} estimated the positive exclusion probability $\mrone$ based on recent history data on the true and false indications; our experimental study in Sec.~\ref{sec:epe:sim} shows that this approach provides a good estimation. 
However, estimating the negative exclusion probability $\mrzero$ using such a history-based approach is challenging due to what we dub {\em the $\mrzero$ vicious cycle problem}: To estimate the negative exclusion probability $\mrzero$, the client should occasionally access the cache upon a negative indication; we dub such accesses {\em speculative accesses}. However, before $\mrzero$ is estimated, the client has no incentive to perform such speculative accesses. \salsa\ breaks the $\mrzero$ vicious cycle using a combined history and learning approach, which we detail in Sec.~\ref{sec:epe:mrzero}.

\paragraph*{Cache Selection}
The work~\cite{Accs_Strategies_ToN} developed several polynomial-time approximation cache-selection algorithms. The study~\cite{chen2020sequential} generalized the cache selection problem and formulated it as a problem of accessing unreliable resources. Furthermore,~\cite{chen2020sequential} proved that this is an NP-hard problem, and suggested approximation algorithms for it. 
However, the model used in~\cite{chen2020sequential} does not allow scaling the miss penalty $\missp$, which is a fundamental part of the cache selection problem. 
Our work will consider the algorithms proposed in~\cite{Accs_Strategies_ToN, FN_aware_ICDCS} as benchmarks.

\subsubsection*{Cache-Content Advertisement}

The work~\cite{fpr_fnr_in_dist_replicas} suggested advertising a fixed-size fresh indicator each time either the estimated false-positive probability or false-negative probability crosses predefined thresholds. This solution ignores bandwidth considerations and, therefore, may result in excessive bandwidth consumption. 
The only work known to us that combines a dynamically scaled indicator and update interval is CAB~\cite{CAB}.
However, CAB considers a single-cache scenario. Consequently,  
CAB's cost model assumes that each false-positive event results in an unnecessary cache access and each false-negative event implies an unnecessary memory access. These assumptions do not hold in a multi-cache system, where the algorithm may access several caches in parallel, and retrieve the requested datum from some of them.
Furthermore, CAB always fully utilizes the given ``bandwidth budget'' for advertisements. We argue that many practical scenarios exhibit a diminishing return in the bandwidth-for-performance tradeoff, thus offering a dramatic bandwidth save at the cost of only a marginal increase in the service cost. We expand on that in Sec.~\ref{sec:advertise_alg}.

\section{The Exclusion Probability Estimation}
\label{sec:epe}

In this section, we describe \salsa's exclusion probability estimation scheme. 
We begin by detailing how \salsa\ estimates $\mrone$ and $\mrzero$ and then evaluate the accuracy of these estimations using real-world traces.
\subsection{Estimating the Positive Exclusion Probability $\mrone$}
\label{sec:epe:mrone}

The positive exclusion probability $\mrone$ captures the probability that the requested item is not in the cache, given a positive indication. The main reason for such false-positive indications is the inherent inaccuracy of the indicator, which sacrifices some accuracy for space efficiency~\cite{Survey18}. An additional cause of false-positive indications is {\em staleness} of the indicator. For instance, if an item $x$ was evicted from the cache after the last time the cache advertised an indicator, a query for $x$ results in a false-positive reply. However, as efficient cache policies typically do not evict an item that is likely to be requested shortly, the impact of staleness is small, and $\mrone$ only moderately increases with the time since the last update~\cite{CAB}.

The work~\cite{Accs_Strategies_Infocom} estimates $\mrone$ using an Exponential Weighted Moving Average (EWMA). However,~\cite{Accs_Strategies_ToN} discards inter-caches dependencies. For instance, consider a multi-cache system, where each missed item is inserted into $k > 1$ caches (e.g., Kaleidoscope~\cite{Kaleidoscope}). In such a system, a single positive indication for the requested item $x$ will likely be a false positive. On the other hand, $k$ positive indications for $x$ tend to be all true positives. To account for this effect, we let the cache keep distinct estimations of $\mrone$ for each possible number of positive indications: $\mrone[0], \dots, \mrone[\numCaches]$. 

In more detail, upon each cache access, the client indicates whether this access is ``regular'' (namely, done upon a positive indication); or ``speculative'' (namely, done upon a negative indication). In addition, the client writes in its request the overall number of positive indications for this item. For each $i = 0, 1, \dots, \numCaches$, the cache counts the number of ``regular'' cache accesses and the number of such accesses that resulted in a miss (false positives). Each time regAccsCnt[$i$] (for any $0 \leq i \leq \numCaches$) reaches some predefined {\em window} size, the cache re-estimates $\mrone[i]$: 
\begin{align}\label{eq:estimate_mrone}
    \mrone[i] \gets \delta_{\mrone} \cdot \frac{\textrm {fpCnt[i]}}{\textrm {regAccsCnt[i]}} + (1 - \delta_{\mrone}) \mrone[i ],
\end{align}
where $0<\delta_{\mrone}<1$ is a smoothness parameter, to be determined later.
After each such re-estimation, the cache resets regAccsCnt[$i$] and fpCnt[$i$].
We assume throughout that the window size is 1/10 of the update interval. Intuitively, this is equivalent to updating an estimation ten times during an update interval. As $\mrone$ is expected to change over time slowly, we set $delta_{\mrone}=0.25$. 
Our simulations in Sec.~\ref{sec:epe:sim} show that these choices well balance the estimated value's responsiveness and smoothness.

\subsection{Estimating the Negative Exclusion Probability $\mrzero$}\label{sec:epe:mrzero}

The negative exclusion probability $\mrzero$ is the probability that the requested item is not in cache $j$, given a negative indication. As there exist multiple indicator structures that are false-negative free~\cite{Survey18}, false-negative events are typically not an inherent by-product of the indicator's structure, but instead result from {\em staleness}: if an item $x$ is inserted into the cache after the last time the cache advertised an indicator, a query for $x$ would result in a false-negative indication. Due to the time-locality of cache workloads, once an item $x$ is inserted into the cache, $x$ will likely be queried again soon. Consequently, $\mrzero$ may rapidly evolve during a trace, depending upon the concrete workload and update interval  ~\cite{CAB, FN_aware_ToN}, 

For each $i = 0, 1, \dots, \numCaches$, 
the cache counts the number of speculative accesses 
and the number of such accesses that are true negative. Then, each time specAccsCnt[$i$] reaches the window size (for any $0 \leq i \leq \numCaches$), the cache re-estimates $\mrzero[i]$: 
\begin{align}\label{eq:estimate_mrzero}
    \mrzero[i] \gets \delta_{\mrzero} \frac{\textrm {tnCnt[i]}}{\textrm {specAccsCnt[i]}} + (1 - \delta_{\mrzero}) \mrzero[i],
\end{align}
where $0<\delta_{\mrzero}<1$ is a smoothness parameter. 
As $\mrzero$ may rapidly evolve over time, we set $\delta_{\mrzero}=0.5$. 
After re-estimating $\mrzero[i]$ using Eq.~\ref{eq:estimate_mrzero}, the cache resets specAccsCnt[$i$] and tnCnt[$i$].
\begin{figure*}
    \centering
    \subfloat[\label{fig:mr1}The estimated positive exclusion probability {$\mrone$} in the Twitter45 (left) and IBM7 (right) traces.]{
        \includegraphics[width=\wideSubfloatWidth]{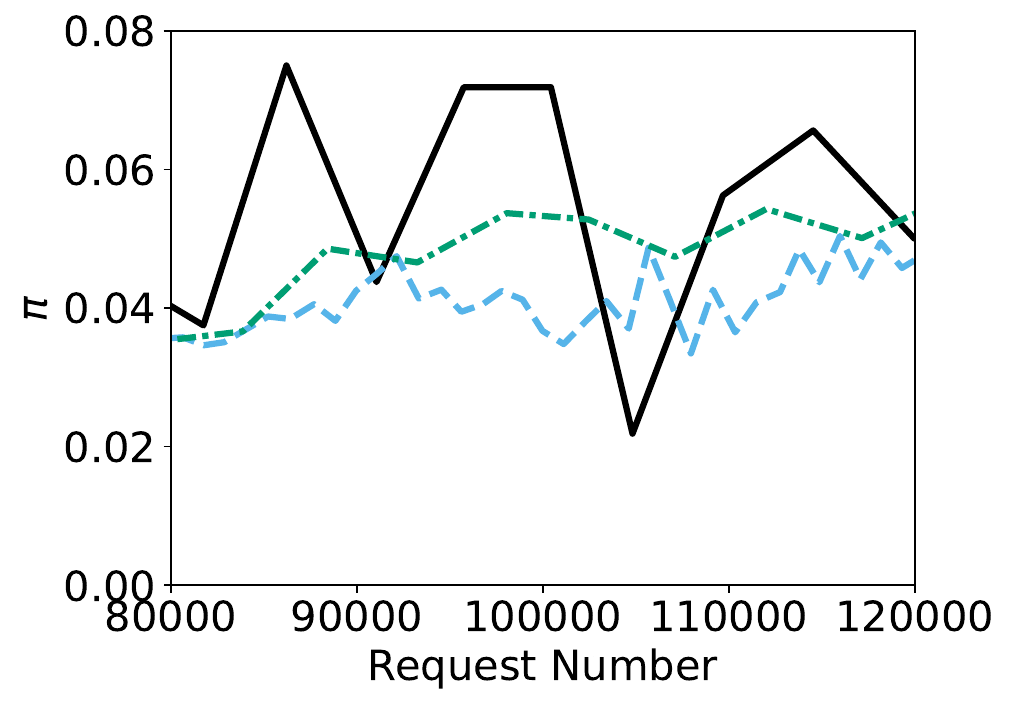}
        \includegraphics[width=\wideSubfloatWidth]{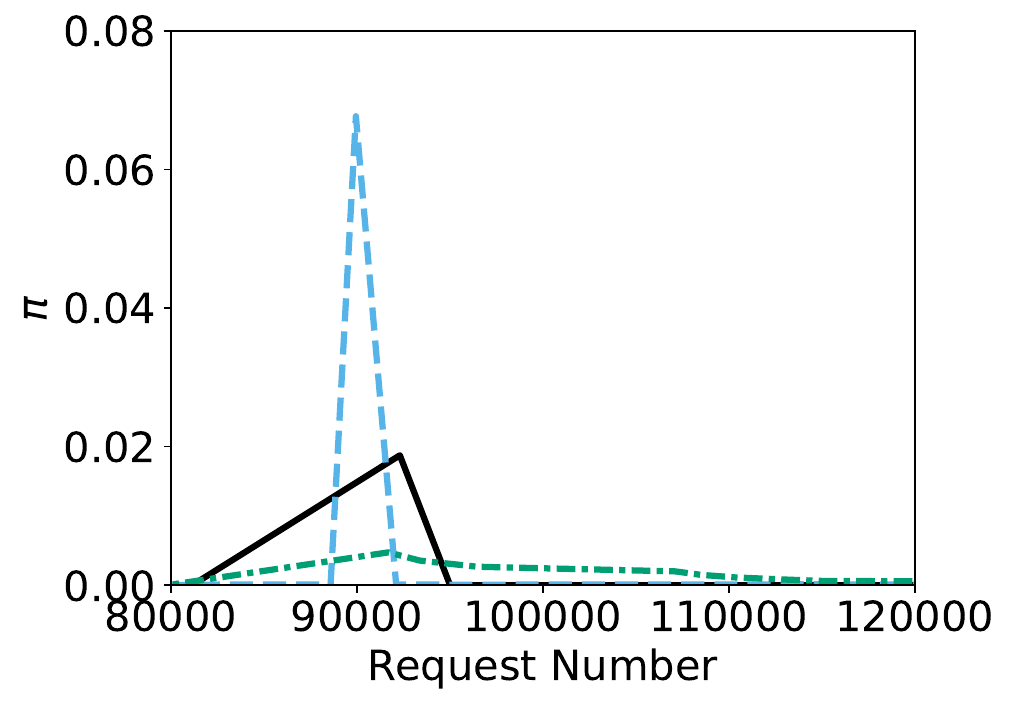}
    }

    \captionsetup[subfloat]{labelformat=empty}
    \subfloat[]{
    \includegraphics{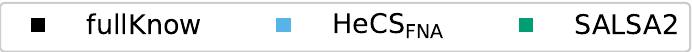}
    }

    \captionsetup[subfloat]{labelformat=parens}
    \subfloat[\label{fig:mr0}The estimated negative exclusion probability {$\mrzero$} in the Twitter45 (left) and IBM7 (right) traces.]{
        \includegraphics[width=\wideSubfloatWidth]{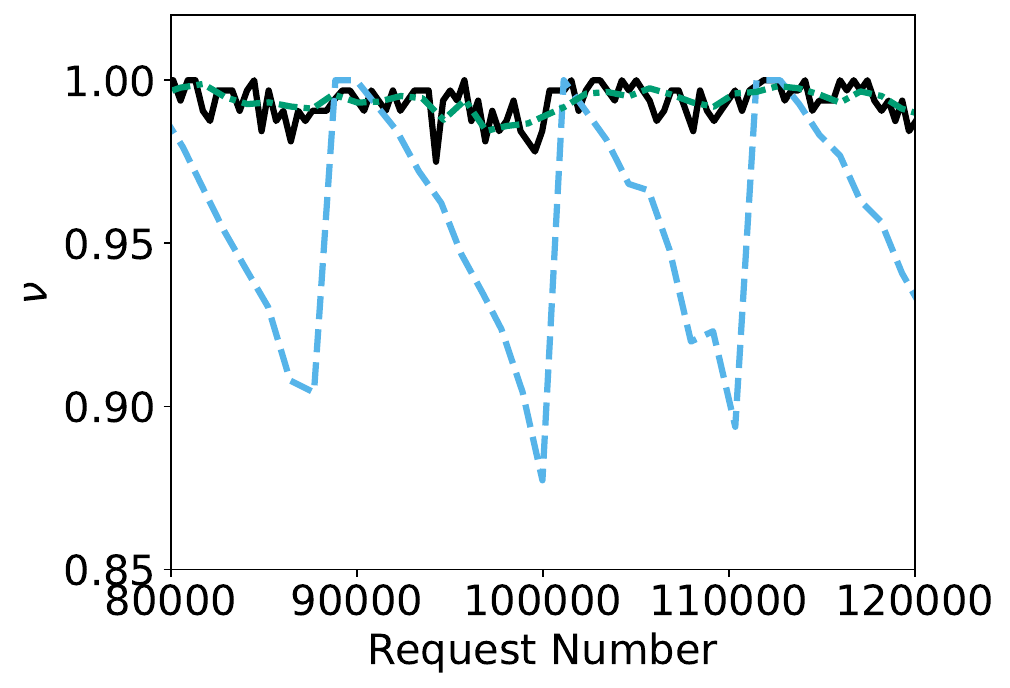}
        \includegraphics[width=\wideSubfloatWidth]{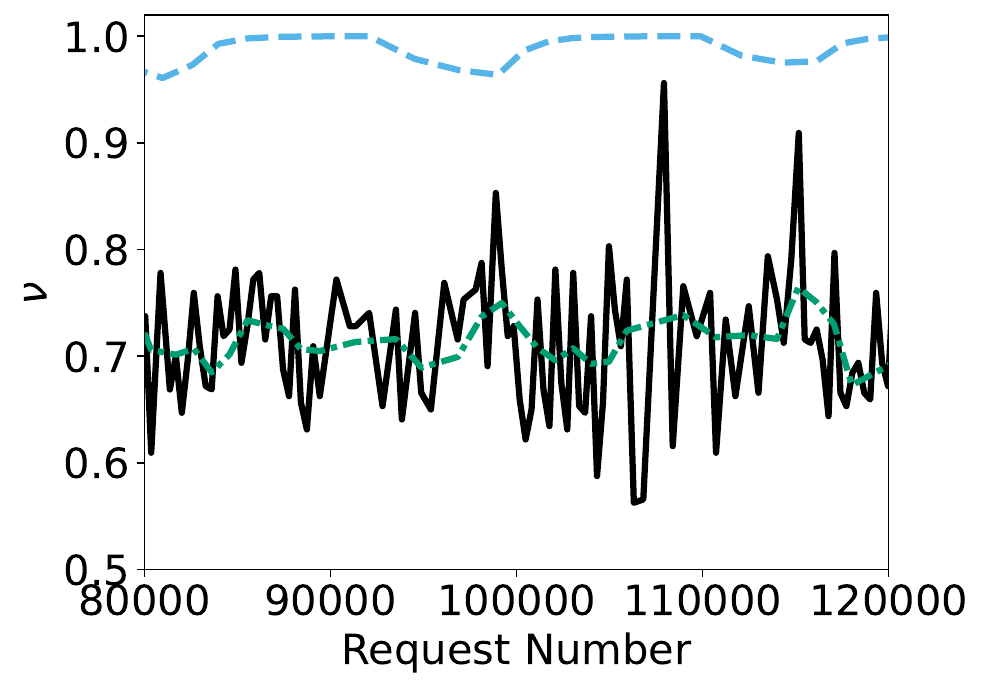}
    }
    \caption{\label{fig:mr}The exact instantaneous exclusion probabilities (fullKnow) and the estimations produced by \pgmfna~\cite{FN_aware_ToN} and \salsa. 
    }
\end{figure*}

\paragraph*{Breaking the $\mrzero$ vicious cycle} As we explained in Sec.~\ref{sec:related_work:taxonomy}, estimating $\mrzero$ using the history-based mechanism described above faces the $\mrzero$ vicious cycle problem: to gather relevant history, the client must perform speculative indications. However, before $\mrzero$ is estimated, the client is not incentivized to access the cache speculatively.

\noindent To break this vicious cycle, \salsa\ initializes, for every $i$, $\mrzero[i]=\mrzeroInit$, where $\mrzeroInit$ is low enough to incentivize the client to perform speculative accesses occasionally. As the miss penalty is typically at least an order of magnitude higher than the cost of cache access, we recommend assigning $\mrzeroInit$ below 0.9. 
If during a trace, $\mrzeroInit$ gets again too high, $\mrzero$ vicious cycle may be closed again, restraining the option to reveal future changes in the workload. 
To tackle that, once in some predetermined epoch, \salsa\ clamps the estimated negative exclusion probability $\mrzero$ to, at most, $\mrzeroInit$. This is done by assigning for every $1 \leq i \leq \numCaches$
\begin{align}\label{eq:mrzeroInit}
    \mrzero[i] \gets \min \set{ \mrzero[i], \mrzero_{\textrm init}}.
\end{align}

To minimize the overhead of these (possibly wasteful) exploration speculative accesses, \salsa\ clamps $\mrzero$ as defined in~\eqref{eq:mrzeroInit} once in an epoch of $10 \cdot \uInterval$, where $\uInterval$ is the update interval. 
Recalling that estimation window is $0.1 \cdot \uInterval$, this can be seen as assigning a single ``learning period'' of $0.1 \cdot \uInterval$, followed by a $9.9 \cdot \uInterval$-length ``inference period''. This way, we bound the overhead of the learning periods to at most 1\% of the overall access cost.

\subsection{Evaluation of the Exclusion Probabilities Estimations}\label{sec:epe:sim}
We now evaluate the accuracy of \salsa's exclusion probabilities estimation. Our testbed considers a system with three caches and two real-world cache traces, Twitter45 and IBM7 (detailed in Sec.~\ref{sec:sim_settings}). 
Each missed item is inserted into a single cache, chosen by a hash of the item's key.
The capacity of each cache is 16K items. Each cache advertises a fresh Bloom filter indicator once in 3200 insertions of new items. For each item in the trace, the client queries all the indicators and accesses all the caches with positive indications (if such exist), and a single cache which is picked u.a.r. from the caches with negative indications.
We consider the following benchmarks:
\begin{inparaenum}[(i)]
    \item The estimations of $\mrone$ and $\mrzero$ performed by \pgmfna~\cite{FN_aware_ToN}. 
    \item {\em fullKnow}: a hypothetical algorithm that always knows whether a desired item was found in the cache (even if the cache was not accessed) and calculates the exclusion probabilities accordingly once in every 320 requests.
\end{inparaenum}

\pgmfna\ does not consider inter-caches dependency; hence, at each point in time, it provides a single estimation of $\mrone$ and a single estimation of $\mrzero$.
fullKnow and \salsa, on the other hand, consider the exclusion probabilities per each possible number of positive indications.
Hence, fullKnow and \salsa\ let each cache maintain eight estimations: $\mrzero[0], \mrzero[1], \mrzero[2], \mrzero[3], \mrone[0], \mrone[1], \mrone[2], \mrone[3]$. 
Out of these values, We plot $\mrzero[0]$ and $\mrone[1]$. 
We focus on these values because, in our testbed, a missed item is inserted into a single cache. As a result, when there are no 
positive indications, there is some probability of a false-negative event in one of the indicators; this case is captured by $\mrzero[0]$. When there is a single positive indication, there exists some probability that it is a false-positive; this case is captured by $\mrone[1]$. Other cases are less of interest, as the result is known a-priori. For instance, when there are two positive indications, one knows a-priori that at least one of them is false-positive. 

Fig.~\ref{fig:mr} depicts the results of our experiments. 
Both
\pgmfna\ and \salsa\ provide good estimations of $\mrone$. 
However, \salsa's $\mrzero$ estimations are significantly more accurate than those of \pgmfna.
Interestingly, \pgmfna's estimation of $\mrzero$ makes a tooth saw pattern, capturing the fact that \pgmfna\ bases its estimation on the numbers of bits set/reset in the indicator since the list advertisement, which typically constantly increase after each advertisement.

\section{The Cache Selection Algorithm}\label{sec:cs}
We now detail \salsa's cache selection algorithm.

Each time a cache updates either $\mrzero[i]$ 
or $\mrone[i]$ for some $i$, the cache sends this update to all the clients. The communication overhead is negligible, as it requires sending only a small integer ($i$) and a single float value ($\mrzero[i]$ or $\mrone[i]$). Furthermore, these periodic updates may be piggybacked on a cache reply.

Alg.~\ref{alg:DC} illustrates how the client selects which caches to access. Upon every request for datum $x$, the client obtains the for each cache $j$ the respective indication $\ind_j(x)$. Recall that $\numPosInds$ denotes the number of positive indications. For each cache, the client assigns the miss probability $\mr_j$ as either the corresponding positive exclusion probability $\mronej[\numPosInds]$ (if the indication is positive) or the negative exclusion probability $\mrzeroj[\numPosInds]$ (else). 
Then, the client calls a cache selection algorithm -- e.g., one of the algorithms in~\cite{Accs_Strategies_ToN}. 
Importantly, by assigning to $\rho_j$ the concrete exclusion probability corresponding to the number of positive indications for the requested datum ($\numPosInds$), \salsa\ captures inter-cache dependencies. 
For instance, in a multi-cache system where a missed item is inserted into several others, this approach ensures a more precise estimation of exclusion probabilities, reducing unnecessary cache accesses. This consideration of inter-cache dependencies is overlooked in state-of-the-art solutions, solutions~\cite{Accs_Strategies_ToN,
FN_aware_ToN, chen2020sequential}.
\begin{algorithm}[t!]
\caption {The Client Cache Selection Algorithm
}
\label{alg:DC} 
\begin{algorithmic}[1]
    \State periodically obtain updated $\mronej, \mrzeroj$ from each cache $j$
    \label{alg:DC:obtain_estimations}
    \For {every request for datum $x$}
        \For {every cache $j$}
    	\If {$\ind_j(x) = 1$}%
             \label{alg:DC:if} 
                \State {$\mrj \gets \mronej[\numPosInds]$}            
            \Else
                \label{alg:DC:else} 
                 \State {$\mrj \gets \mrzeroj[\numPosInds]$}
            \EndIf
             \label{alg:DC:if_end} 
            \State $D = \textrm{cacheSelectionAlg}(\setCaches, \vec{\accsCost}, \vec{\mr})$
        \label{alg:DC:call_PGM}
        \State access $D$%
        \label{alg:access_D}
        \EndFor
    \EndFor
\end{algorithmic}
\end{algorithm}

\section{The Indicator Advertisement Algorithm}
\label{sec:advertise_alg}

In this section, we describe \salsa's advertisement mechanism. We begin with a high-level description and then detail \salsa's two modes of operations, namely full-indicator mode and delta mode.

\subsection{High-level Description}

\salsa\ aims to minimize the communication overhead by considering a given bandwidth budget of $\bwBudget$ advertised bits per new item's insertion. Importantly, this is viewed as a soft constraint: on the one hand, \salsa's communication overhead may instantaneously exceed $\bwBudget$; on the other hand, when the setting and current workload allow, \salsa\ utilizes a significantly lower bandwidth.

Initially, \salsa\ operates in the full indicator mode, where the bandwidth constraint is enforced by advertising an $\indSize$-bits indicator once in at least $\frac{\indSize}{\bwBudget}$ insertions. To boost performance, \salsa\ dynamically scales the indicator size according to the perceived exclusion probabilities. 
The scaling mechanism operates as follows. 
As explained in Sec.~\ref{sec:epe:mrone}, false positives occur mainly due to an inaccurate indicator. Hence, if 
the positive exclusion probability exceeds some threshold $\mrone^{th}$, \salsa\ up-scales the indicator.
Alternatively, a low negative exclusion probability indicates multiple recent false negative events, due to staleness (recall Sec.~\ref{sec:epe:mrzero}). To mitigate staleness, one should advertise fresh indicators more frequently (and down-scale the indicator accordingly to comply with the bandwidth budget).
After each such indicator scaling, \salsa\ sets the update interval to $\uInterval = \frac{\indSize}{\bwBudget}$, where $\indSize$ is the new indicator size.

Before advertising a full indicator, \salsa\ checks whether transmitting a delta update can save bandwidth. If transmitting a delta update is cheaper, the algorithm switches to delta mode. 

While in delta mode, \salsa\ 
keeps a minimal update interval to reduce staleness, and regulates the communication overhead by scaling the indicator size. 
However, if the indicator's content changes too fast, advertising delta updates, which list all the flipped bits since the previous update, may 
require too much bandwidth. 
In such a case, \salsa\ returns to full-indicator mode.

\subsection{Full-indicator Mode}

In a distributed caching system, a cache may experience multiple accesses before a new item is inserted, or vice versa, namely, multiple insertions of new items between two successive accesses.
Hence, while in full indicator mode, \salsa\ is called upon an event of either a cache access or a new item insertion. 

\begin{algorithm}[t!]
\caption {After every cache access (full indicator mode)
}
\label{alg:upon_cache_accs} 
\begin{algorithmic}[1]
    \If {insCnt $ > \uInterval$}
        \If {$\mrone[\numPosInds] > \mrone^{th}$}
            \State ScaleUp ()
            \State advertiseFullIndicator()
        \ElsIf {$\mrzero[\numPosInds] < \mrzero^{th}$}
            \State ScaleDown ()
            \State advertiseFullIndicator()
        \EndIf
    \EndIf
\end{algorithmic}
\end{algorithm}

Upon every cache access, \salsa\ estimates the exclusion probabilities as detailed in Sec.~\ref{sec:epe}, and then runs the pseudo-code detailed in Alg.~\ref{alg:upon_cache_accs}. 
The algorithm uses the following procedures: 
\begin{itemize}
    \item {\em ScaleUp()}: scale the indicator to $\indSize = \min\set{1.1 \cdot \indSize, \indMaxSize}$, and then adapt the update interval: $\uInterval = \frac{\indSize}{\bwBudget}$. 
    \item {\em ScaleDown()}: scale the indicator to $\indSize = \max\set{\frac{\indSize}{1.1}, \indMinSize}$, and then assign $\uInterval = \frac{\indSize}{\bwBudget}$. 
    \item {\em advertiseFullIndicator()}: advertise a full indicator and then resets insCnt, which counts the number of insertions since the last advertisement. Reset also tnCnt[$i$] and specAccsCnt[$i$]=0 for every $i$, as these counters correspond to the indicator' staleness, which is nullified upon advertisement.
\end{itemize}

After the insertion of a new item, \salsa\ runs the algorithm illustrated in Alg.~\ref{alg:after_insertion:full}. 
The procedure xmtDeltaIsCheaper() returns true iff switching to delta mode may save bandwidth; xmtDeltaIsCheaper()'s implementation is detailed in the appendix. If this is beneficial, \salsa\ switches to delta mode (ln.~\ref{alg:after_insertion:full:if_delta_is_cheaper}-\ref{alg:after_insertion:full:if_delta_is_cheaper_end}). Else, \salsa\ stays in full-indicator mode. 
To reduce the computational overhead, \salsa\ calls xmtDeltaIsCheaper() only when insCnt's value is precisely $\uInterval$.
There is no need to run the procedure for larger insCnt values, because when more items are inserted into the cache, there are more changes to report. Accordingly, the benefit of switching to delta mode can only decrease.


To save bandwidth, \salsa\ tries to delay the advertisement even after $\uInterval$ insertions since the last update. However, a cache that does not advertise its content for too long may be underutilized. Hence, 
ln.~\ref{alg:after_insertion:full:clamp_if}-\ref{alg:after_insertion:full:clamp_if_end} clamp
the maximum number of insertions between subsequent indicator's advertisements to $\mathcal{C}$ times the update interval $\uInterval$, where $\mathcal{C}$ is a small integer.

\subsection{Delta Mode}

Delta updates carry the promise to decrease costs significantly. However, this potential cost reduction comes with a significant drawback, 
namely, high sensitivity to lost/corrupted packets: 
A single lost delta update results in the user having a wrong view of the cache's content {\em forever}.

To tackle this problem, we introduce a {\em synchronization period}. In addition to advertising delta updates, the cache advertises a full indicator once in a synchronization period of $R \cdot \uInterval$ insertions. The parameter $R$, which regulates the frequency of the synchronization advertisements, reflects the reliability of the communication. If the channel is highly unreliable, one should set $R=1$, implying that the bandwidth needed in delta mode consists of the overall size of the delta updates sent during the last $\uInterval$ insertions plus a single full indicator. Subsequently, 
advertising delta mode requires more bandwidth than merely advertising a full indicator -- namely, keeping in full-indicator mode. As a result, if $R=1$, the algorithm never enters the delta mode. Conversely, if the communication is fully reliable, one may set $R=\infty$, implying that no synchronization advertisements are needed, thus increasing the incentive to use delta mode.
The periodical synchronization advertisement also allows us to scale the indicator, as we detail shortly.

The scheme that \salsa\ runs after each item insertion while in delta mode is depicted in Alg.~\ref{alg:after_insertion:delta}. 
The procedure estimatedBw() estimates the BW consumption, given the indicator size; we detail estimatedBw() in the appendix.
The algorithm first checks whether it's time for a synchronization advertisement. If the answer is positive, the algorithm checks whether it can satisfy the BW constraint while staying in delta mode (ln.~\ref{alg:after_insertion:delta:if_can_stay_delta}). If the answer is positive again, the algorithm scales the indicator to be as close as possible to fully utilizing the bandwidth budget $\bwBudget$, as doing so may provide both good performance (thanks to using an as-large-as-possible indicator) and low communication overhead (due to staying in delta mode). 
This optimized indicator size $\indSize^j$ is picked out of a list $\mathcal{I}$ of optional sizes between $\indMinSize$ and $\indMaxSize$.
If the expected BW consumption is above the budget even if the indicator is down-scaled to its minimal size, \salsa\ switches back to full mode (ln.~\ref{alg:after_insertion:delta:else_can_stay_delta}-\ref{alg:after_insertion:delta:if_can_stay_delta_end}).

If the synchronization period is not over yet but some $\uIntervalMinFeasible$ insertions have passed since the last delta update, the algorithm sends a delta update (ln.~\ref{alg:after_insertion:delta:if_need_to_send_delta}-\ref{alg:after_insertion:delta:send_delta}).

\begin{algorithm}[t!]
\caption {After every item insertion (full indicator mode)}
\label{alg:after_insertion:full} 
\begin{algorithmic}[1]
    \If {insCnt = \uInterval \ \AND xmtDeltaIsCheaper()} \label{alg:after_insertion:full:if_delta_is_cheaper} 
        \State mode $\gets$ delta
        \label{alg:after_insertion:switch_to_delta} 
        \State $\uInterval \gets \uInterval_{\textrm min}$
        \State advertise delta update and reset insCnt
    \label{alg:after_insertion:full:if_delta_is_cheaper_end} 
    \ElsIf {insCnt $ > \mathcal{C} \cdot  \uInterval$}
        \label{alg:after_insertion:full:clamp_if}
        \State advertiseFullIndicator()
    \EndIf
    \label{alg:after_insertion:full:clamp_if_end}
\end{algorithmic}
\end{algorithm}

\begin{algorithm}[t!]
\caption {After every item insertion (delta mode)}
\label{alg:after_insertion:delta} 
\begin{algorithmic}[1]
    \If {\# insertions in this period = $\period \cdot \uInterval$}
        \If {estimatedBw($\indSize_{min}$) $ \leq B$}
        \label{alg:after_insertion:delta:if_can_stay_delta}
            \State $I \gets \min_{\indSize^j \in \mathcal{I}} \set{\abs{\textrm{estimatedBw}(I_j) - B}}$     
        \Else
            \label{alg:after_insertion:delta:else_can_stay_delta}
            \State mode $\gets$ Full
            \State $\uInterval \gets \floor[\Big]{\frac{\indSize}{\bwBudget}}$
            \label{alg:after_insertion:delta:if_can_stay_delta_end}
        \EndIf
        \State \# insertions in this period $\gets$ 0
        \State advertiseFullIndicator()
    \ElsIf {insCnt mod $\uInterval_{min}=0$}
        \label{alg:after_insertion:delta:if_need_to_send_delta}
        \State send delta-update
        \label{alg:after_insertion:delta:send_delta}
    \EndIf
    \label{alg:df}
\end{algorithmic}
\end{algorithm}

\section{Simulation Study}\label{sec:sim}
In this section, we evaluate \salsa's performance in various scenarios.

\subsection{Simulation Settings}\label{sec:sim_settings}

\textbf{Traces.}
We use the following workloads. 
\begin{inparaenum}[(i)]
\item {\em Wiki}: Requests to Wikipedia pages~\cite{WikiBench}.\footnote{The trace includes requests made on Sep. 22, 2007, from 06:12 and on.}
\item {\em Scarab}:
A trace from Scarab Research, a personalized recommendation system for e-commerce~\cite{Scarab_and_Gradle_traces}. \item {\em F1, F2}: Traces from a financial transaction processing system~\cite{Umass_traces}.
\item {\em IBM1, IBM7}: Traces from the IBM Cloud Object Storage service~\cite{snia}.
\item {\em Twitter17, Twitter45}: Traces collected from Twitter cache clusters~\cite{snia}. 
\end{inparaenum}
We chose these traces because they represent versatile tasks and characteristics. The traces' characteristics are analyzed in 
Tab.~\ref{tab:traces}, which presents the mean and standard deviation between sequencing requests for the same item; the statistic considers only items that were requested at least twice. The table also presents the ratio of singular items, namely, items requested only once along the trace. The table shows that the characteristics of the traces vary highly. In particular, IBM1 and IBM7 exhibit high recency, captured by very short inter-arrival (IBM1) or zero singular items (IBM7). Conversely, very low recency characterizes F1 and F2, which exhibit high inter-arrivals, and Twitter45, in which most requests are singular. To gain some intuition about the impact of the trace's characteristic, consider Fig.~\ref{fig:mr} again. Twitter45's plots exhibit high exclusion probabilities, representing that an item is unlikely to be found in the cache either when the indication is positive or negative. Conversely, IBM7's exclusion probabilities are low, capturing a higher recency, resulting in a higher hit ratio.

\begin{table}[]
    \caption{Characteritics of the traces used for evaluation. An inter-arrival is the number of requests between two sequencing requests for the same item. The ratio of singulars captures the number of requests for items requested only once during the trace over the total number of requests.}\label{tab:traces}
    \centering
    \footnotesize
    \begin{tabular}{|c|c|c|c|}
        \hline
         Trace & mean inter-arrival & stdev. inter-arrival & ratio of singulars\\ \hline
        Wiki		&  59185 	&349031 	&5.15e-05 \\ \hline
        Scarab		& 166828 	&584146 	&4.12e-02 \\ \hline
        F1			& 497177 	&795690 	&2.68e-02 \\ \hline
        F2			& 341961 	&1215530 	&1.61e-02 \\ \hline
        IBM1		& 785 		&5772 		&5.15e-03 \\ \hline
        IBM7		& 155612 	&688369 	&0        \\ \hline
        Twitter17	& 143452 	&680674 	&1.46e-02 \\ \hline
        Twitter45	& 156053 	&287967 	&6.88e-01 \\ \hline
    \end{tabular}
\end{table}

\textbf{Benchmark algorithms.}
We consider the following benchmarks.
\begin{inparaenum}[(i)]
\item {\em \pgmfna}: \pgmfna\ algorithm~\cite{FN_aware_ToN}. 
\item 
{\em  \pif}: a hypothetical ideal strategy that always knows where the requested datum is stored and picks the cheapest option accordingly.
\end{inparaenum}
Both \salsa\ and \pgmfna\ use the PGM cache selection algorithm, which provides approximation guarantees at a polynomial run-time~\cite{Accs_Strategies_ToN}.

\textbf{Caches.}
Our baseline system consists of three caches, with access costs 1, 2, and 3. 
Each cache applies the \emph{Least Recently Used (LRU)} eviction policy. 
We express the size of a cache in terms of the maximal number of elements the cache can store.
We consider a system-wide distribution policy that inserts missed items into one or more caches the controller chooses. Such an approach is common in large distributed systems, such as Memcached~\cite{Memcached} and Kademlia~\cite{Kaleidoscope}, for load balancing and maximizing the cached content. Specifically, we insert each missed item into $\floor*{\frac{\numCaches}{3}}$ caches, chosen by hashing the item's key. In the baseline 3-caches scenario, this corresponds to inserting each item into a single cache. 

\textbf{Indicators.}
To advertise a fresh indicator, cache $j$  generates an updated Simple Bloom Filter (SBF)~\cite{Bloom}. The initial size of the SBF $j$ is $\indSize^{init} = 14 \cdot \cacheSize_j$, where $\cacheSize_j$ is the size of cache $j$. Using the optimized number of hash functions, this translates to a designed false-positive ratio of 0.1\%~\cite{Survey12}.
The initial update interval is $0.1 \cdot \cacheSize_j$. 
SALSA2 sets it bandwidth budget accordingly, namely, $\bwBudget=\frac{\indSize}{\uInterval} = \frac{14 \cdot \cacheSize_j}{0.1 \cdot \cacheSize_j}=140$ bits per insertion.

\noindent Each cache also locally saves the last generated (``stale'') SBF. 
\salsa\ compares the updated and the stale indicator to decide whether to switch to ``delta'' mode (ln.~\ref{alg:after_insertion:full:if_delta_is_cheaper} in Alg.~\ref{alg:after_insertion:full}). \pgmfna\ periodically compares the stale and updated indicators to estimate the exclusion probabilities~\cite{FN_aware_ToN}.
\pgmfna\ always keeps the initial indicator size and update interval~\cite{FN_aware_ToN}, while 
\salsa\ dynamically scales them, as detailed in Sec.~\ref{sec:advertise_alg}.

\textbf{Evaluation metrics.} We consider the mean service cost per request over the entire input. To make a meaningful comparison, we divide the service cost of each algorithm (\pgmfna\ and \salsa) by that of the perfect indicator (\pif) for the same setting.

\noindent We also consider the bandwidth consumption. As \pif\ uses a hypothetical zero-bandwidth oracle for the existence of items in the caches, the bandwidth cannot be normalized. Instead, we consider the bits-per-request, calculated by dividing the overall bit-count of advertisements by the number of requests in the trace. 

\textbf{Other system and algorithm parameters.} 
In our experiments, we will vary the caches' sizes and numbers and the miss penalty to study their effect on performance.
The rest of \salsa's parameters are detailed in Tab.~\ref{tab:sim_params}. 
We performed some experiments to study the impact of these parameters. As our results show that these parameters do not significantly impact the results or provide any important insight into the problem, we omit these results from this work.

\begin{table}
    \caption{\salsa's simulation settings.}
    \label{tab:sim_params}
    \centering
    \begin{tabular}{|c|c|c|c|c|c|c|c|}
    \hline
         $\delta_{\mrone}$ & $\delta_{\mrzero}$ & $\mrone^{th}$ & $\mrzero^{th}$ & $\mrone^{init}$ & $\mrzero^{init}$ & $\mathcal{C}$ & $R$\\ 
    \hline
         0.5 & 0.25 & 0.01 & 0.88 & 0.001 & 0.08 & 2 & 10\\ 
    \hline
    \end{tabular}
\end{table}

\subsection{Impact of Miss Penalty and Cache Size}

\begin{figure*}
    \captionsetup[subfloat]{labelformat=empty}
    \centering
    \subfloat[Cache Size = 4K, miss penalty=30]{
        \includegraphics[width=\subfloatWidth]{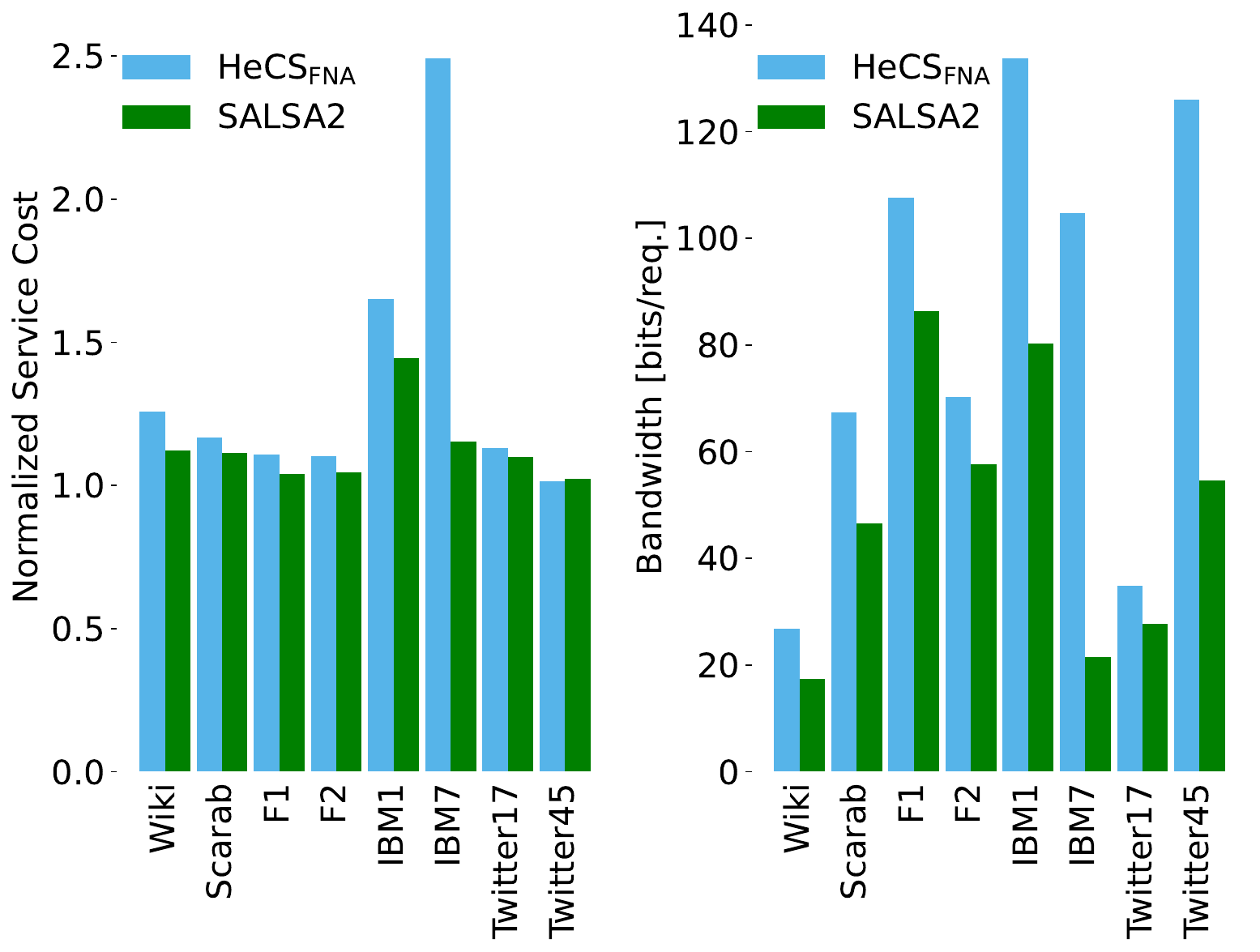}
    }
    \subfloat[Cache Size = 4K, miss penalty=10]{
        \includegraphics[width=\subfloatWidth]{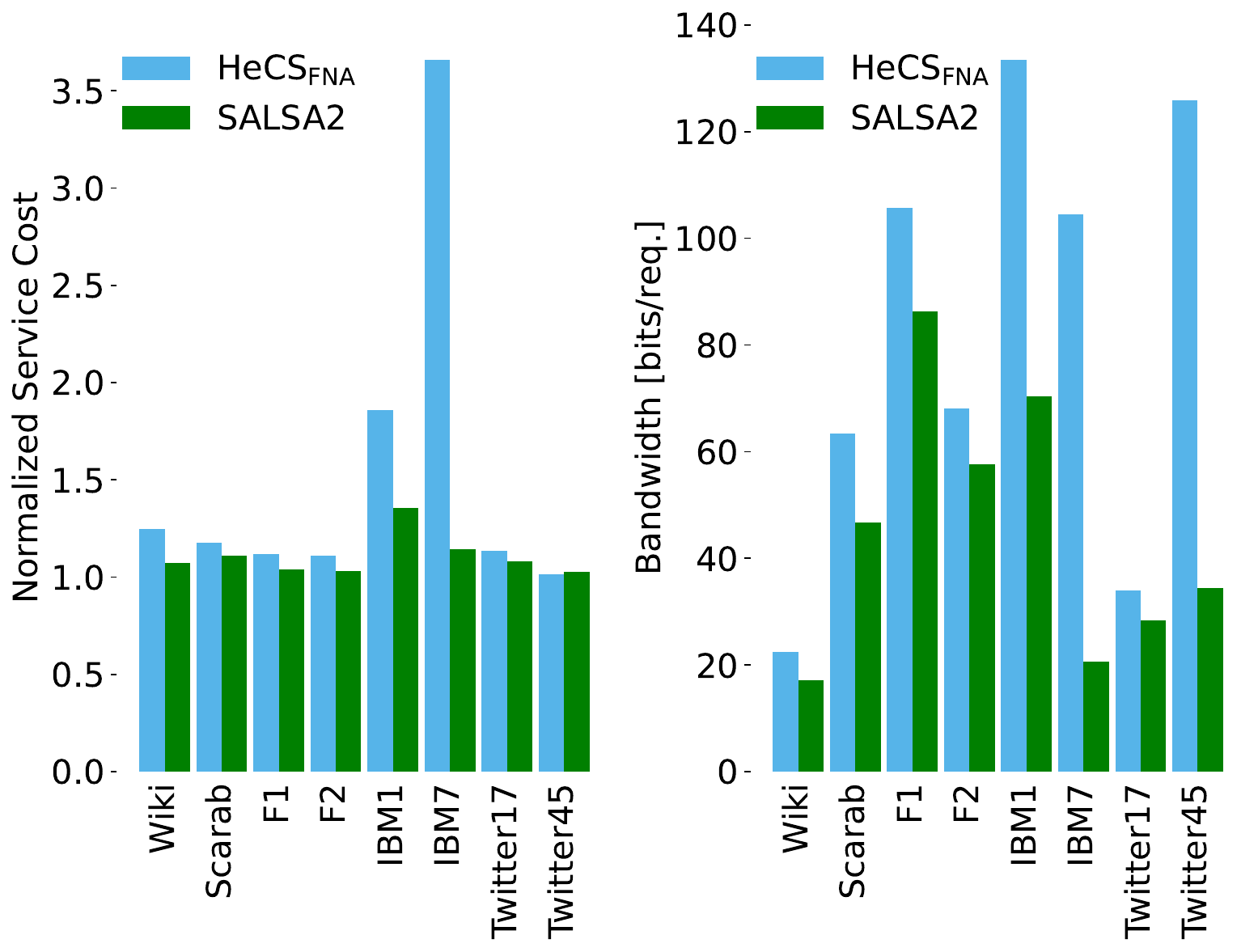}
    }
    \subfloat[Cache Size = 4K, miss penalty=300]{
        \includegraphics[width=\subfloatWidth]{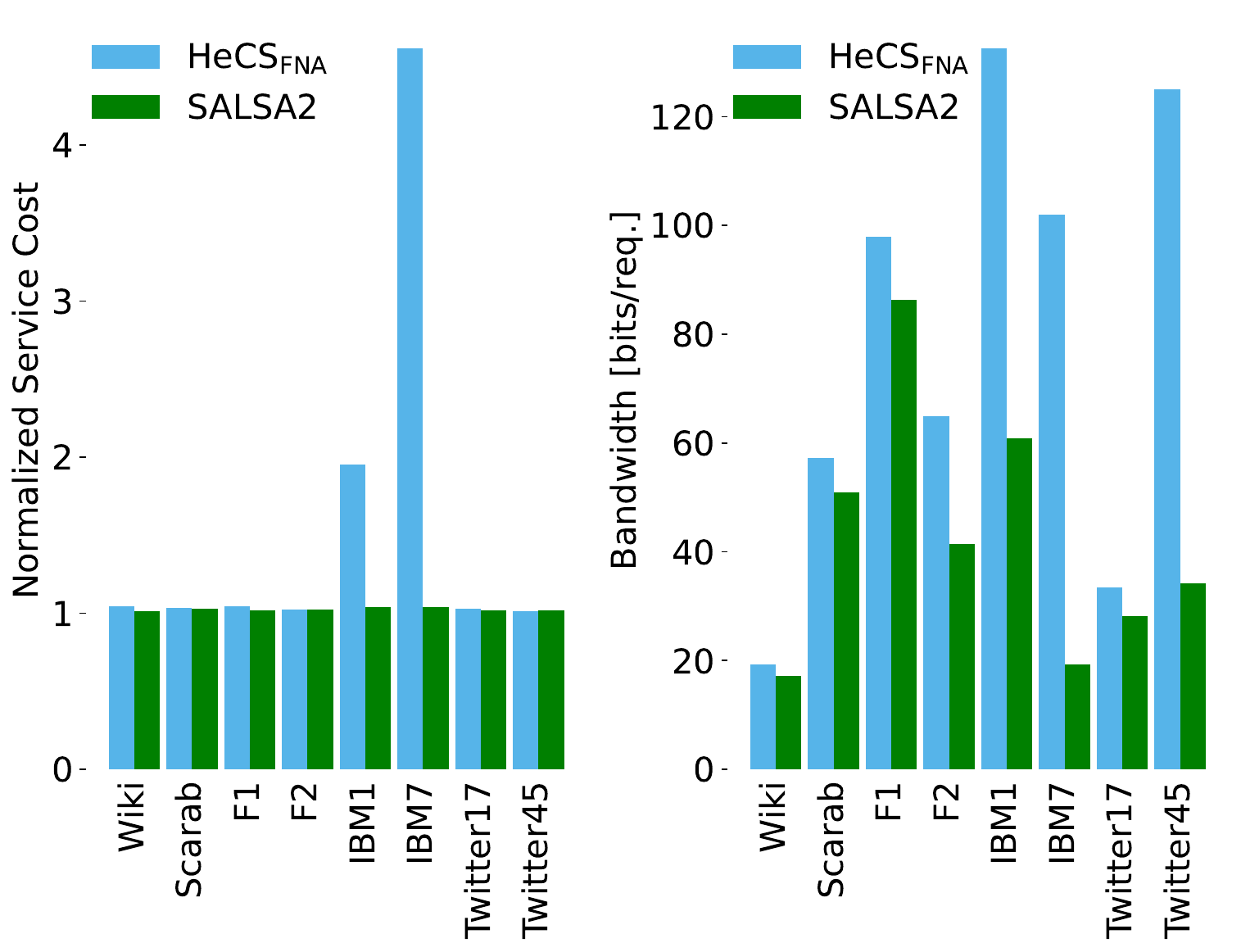}
    }

    \subfloat[Cache Size = 16K, miss penalty=10]{
        \includegraphics[width=\subfloatWidth]{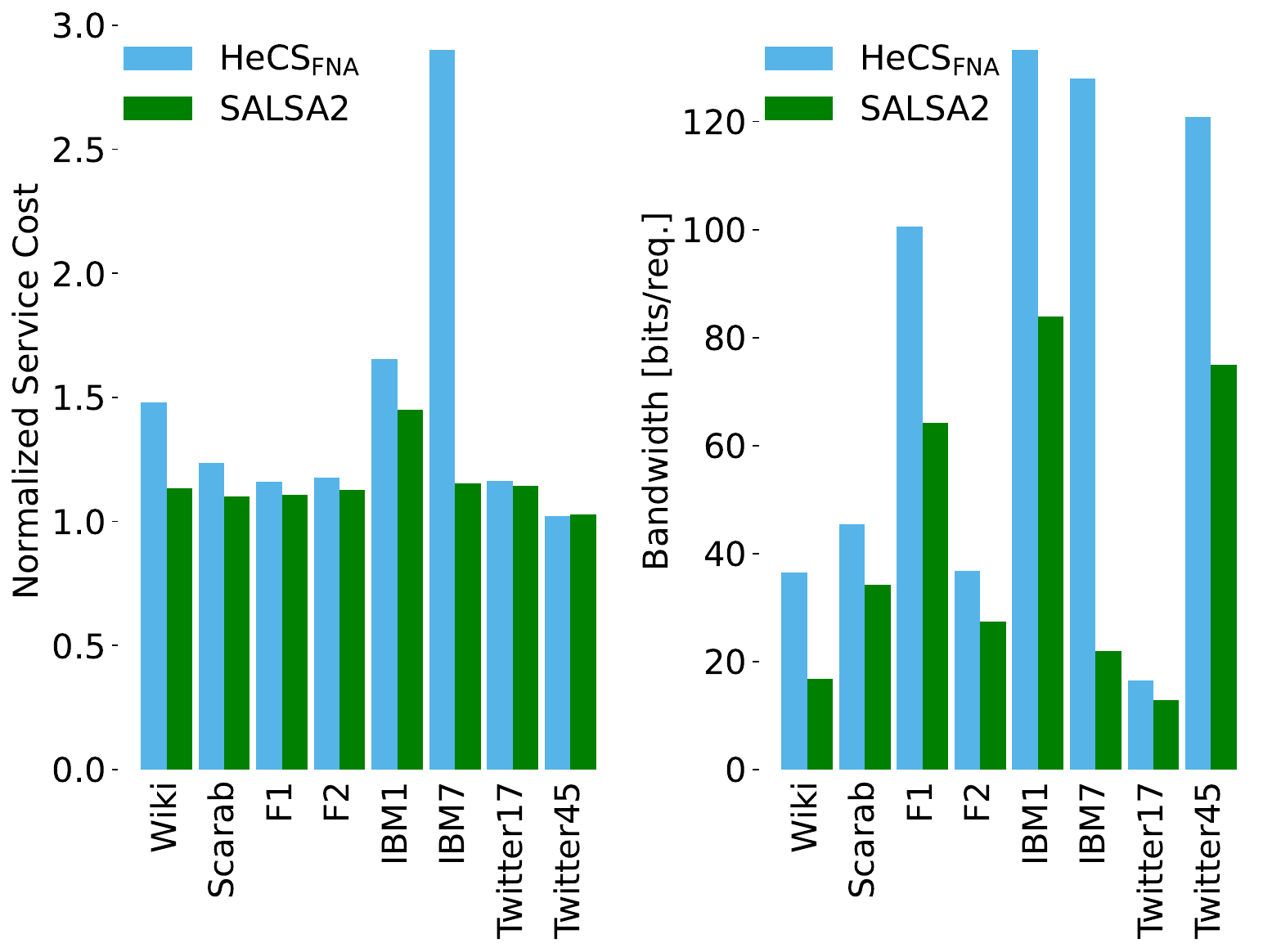}
    }
    \subfloat[Cache Size = 16K, miss penalty=30]{
        \includegraphics[width=\subfloatWidth]{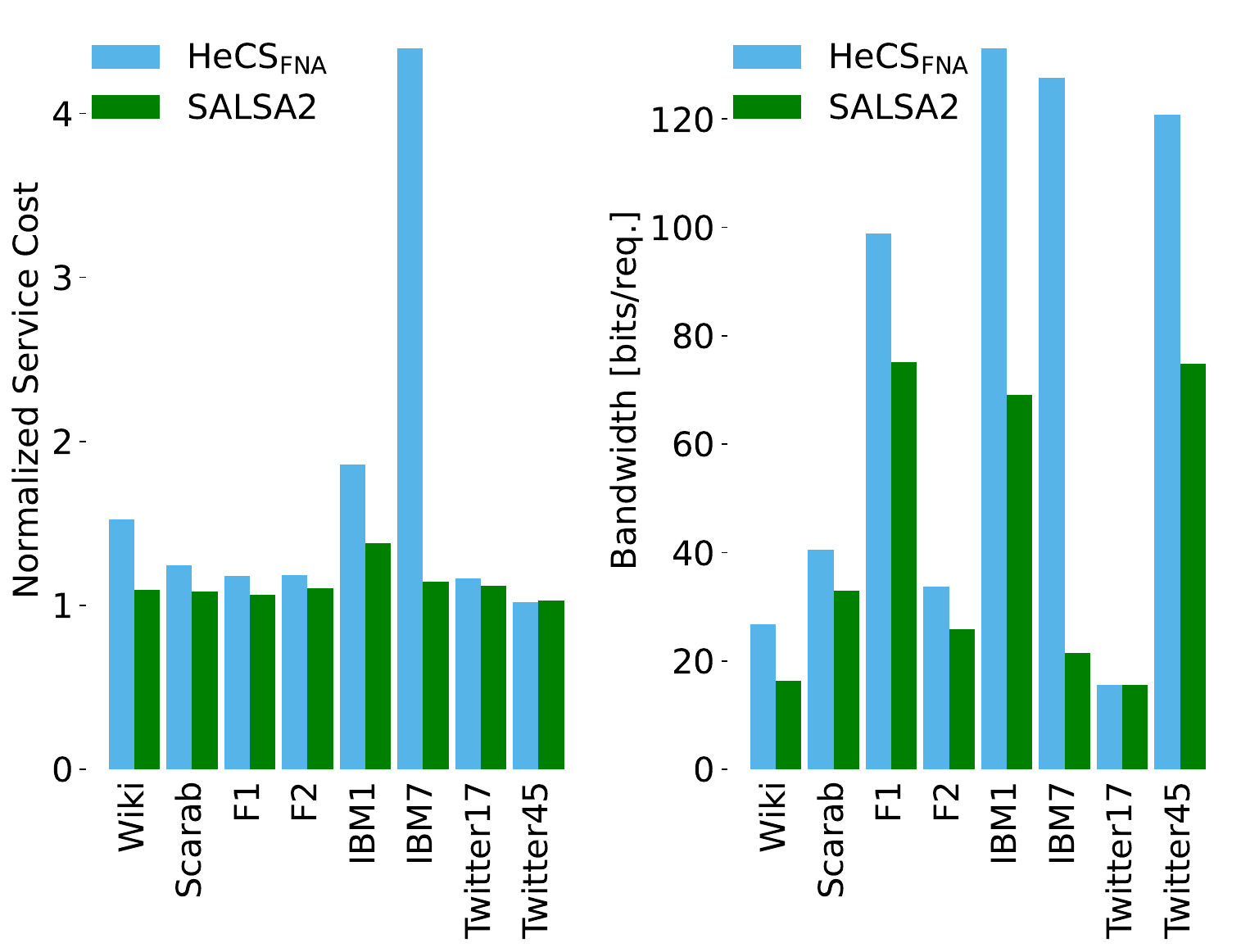}
    }
    \subfloat[Cache Size = 16K, miss penalty=300]{
        \includegraphics[width=\subfloatWidth]{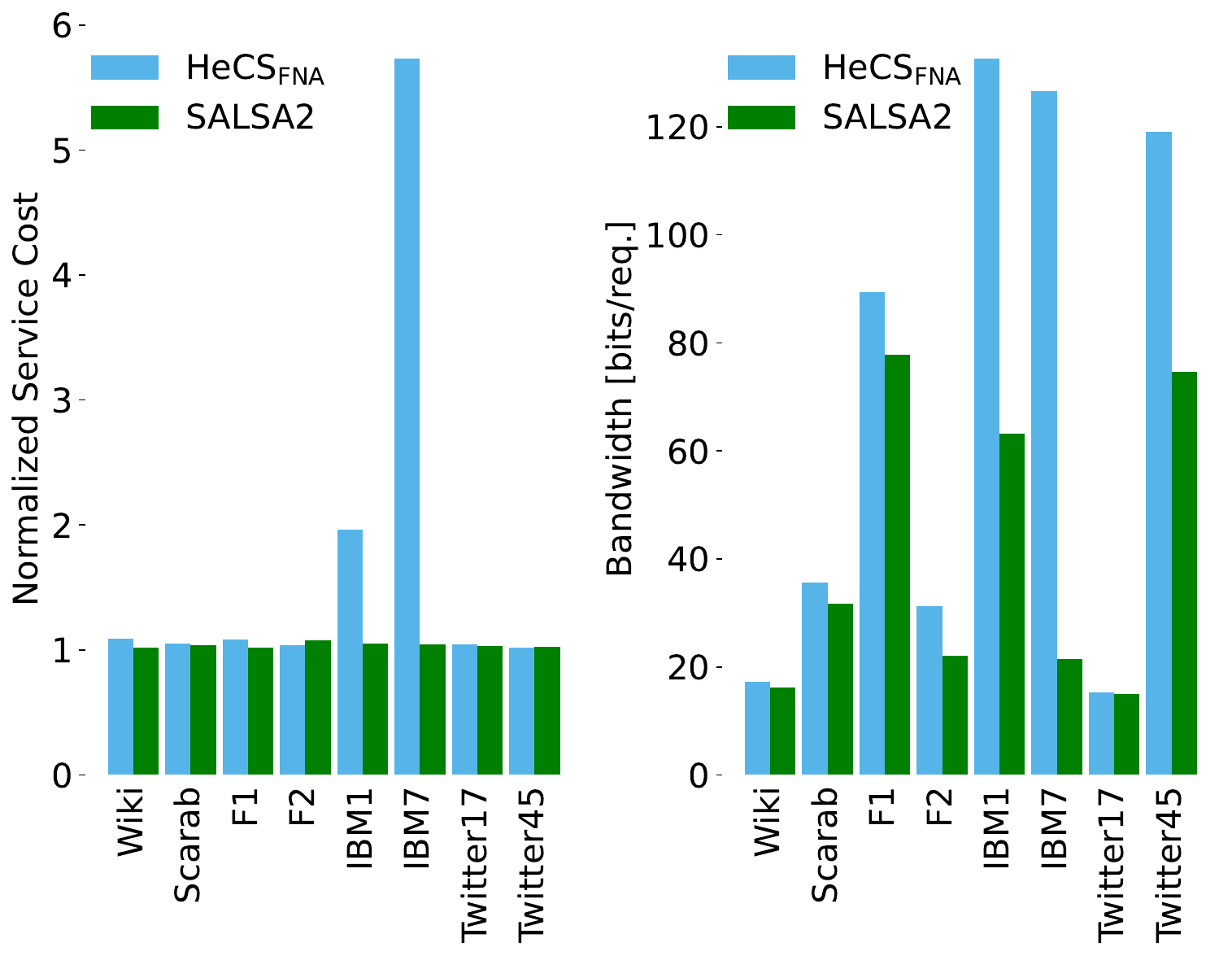}
    }

    \subfloat[Cache Size = 64K, miss penalty=10]{
        \includegraphics[width=\subfloatWidth]{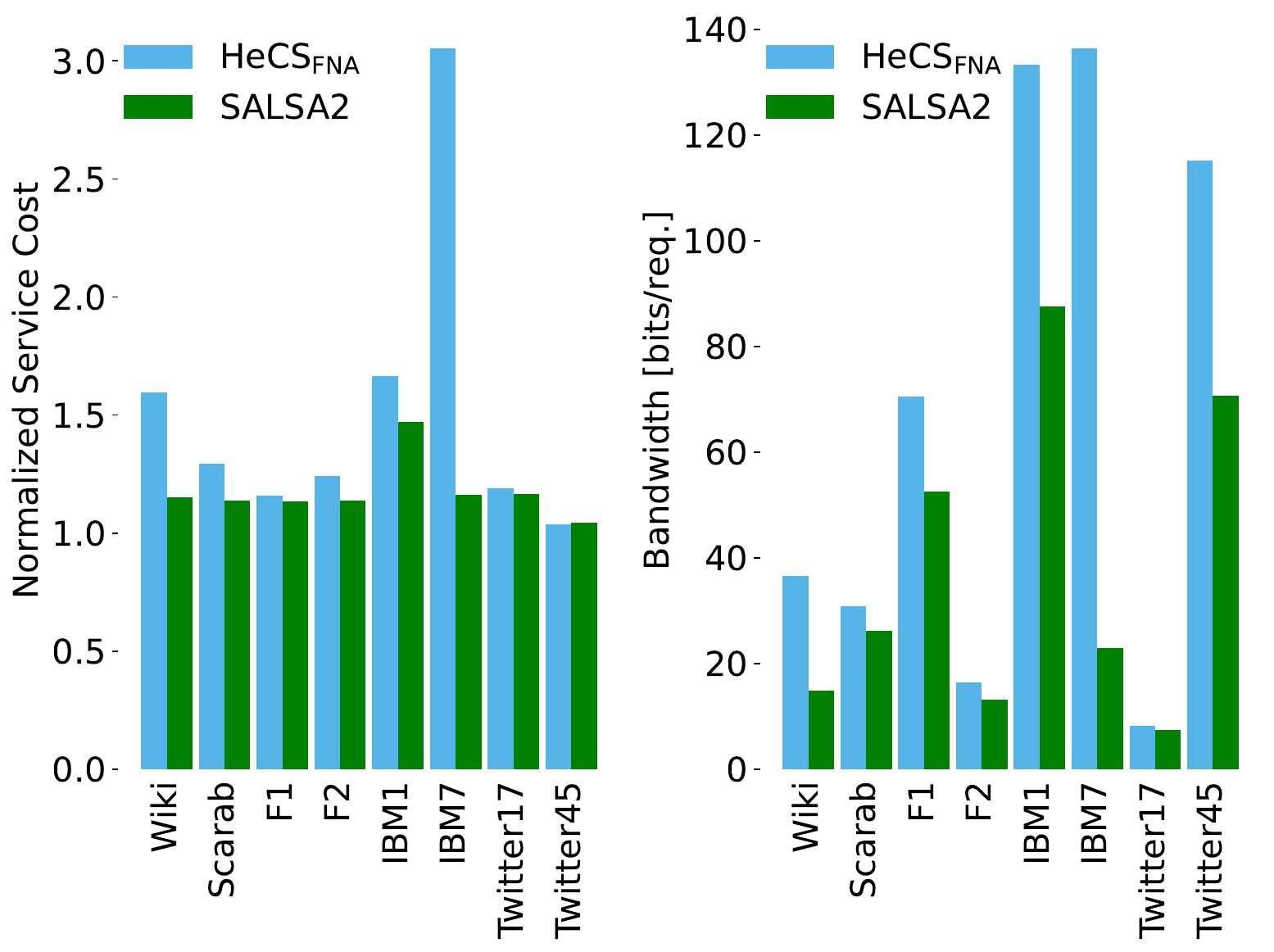}
    }
    \subfloat[Cache Size = 64K, miss penalty=30]{
        \includegraphics[width=\subfloatWidth]{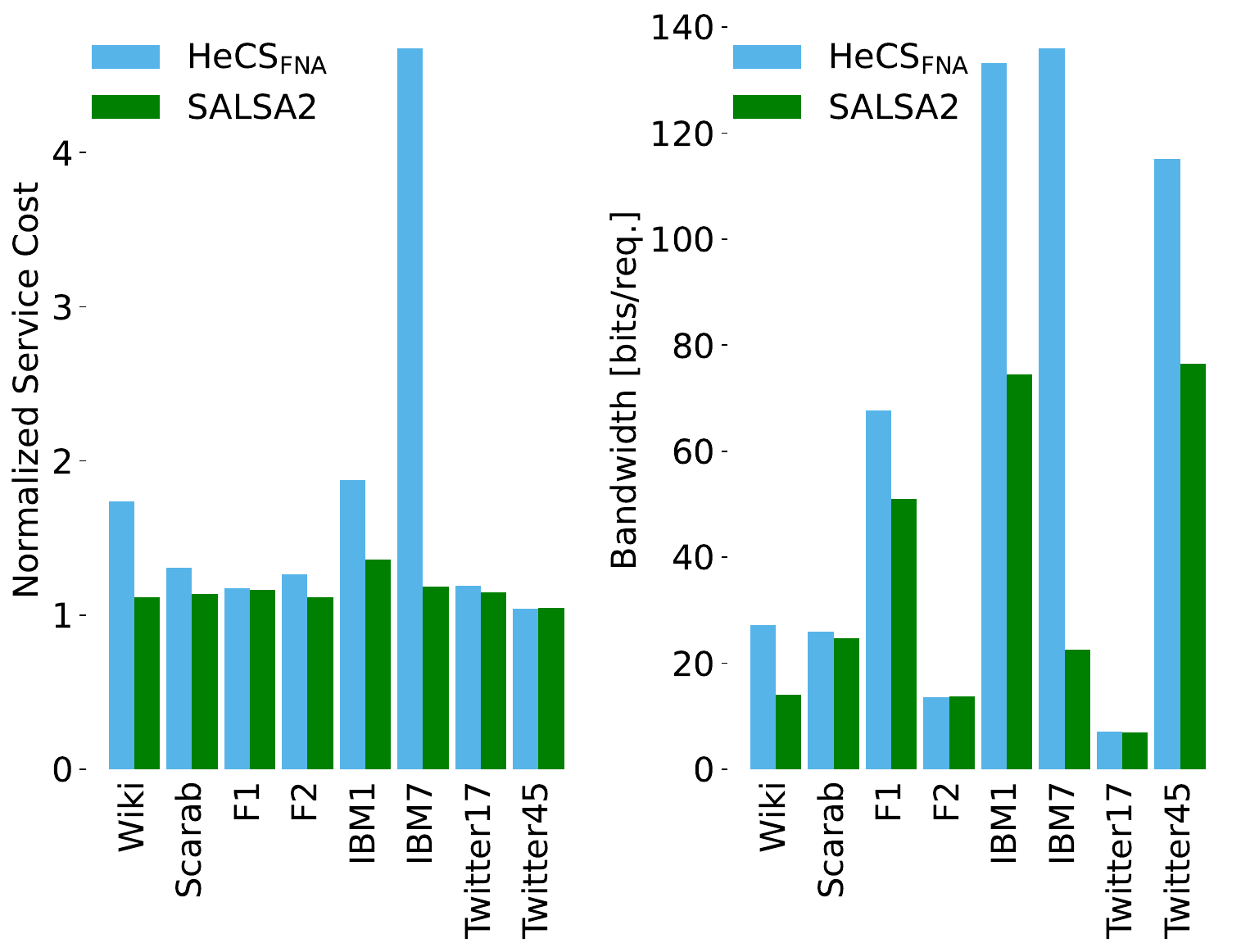}
    }
    \subfloat[Cache Size = 64K, miss penalty=300]{
        \includegraphics[width=\subfloatWidth]{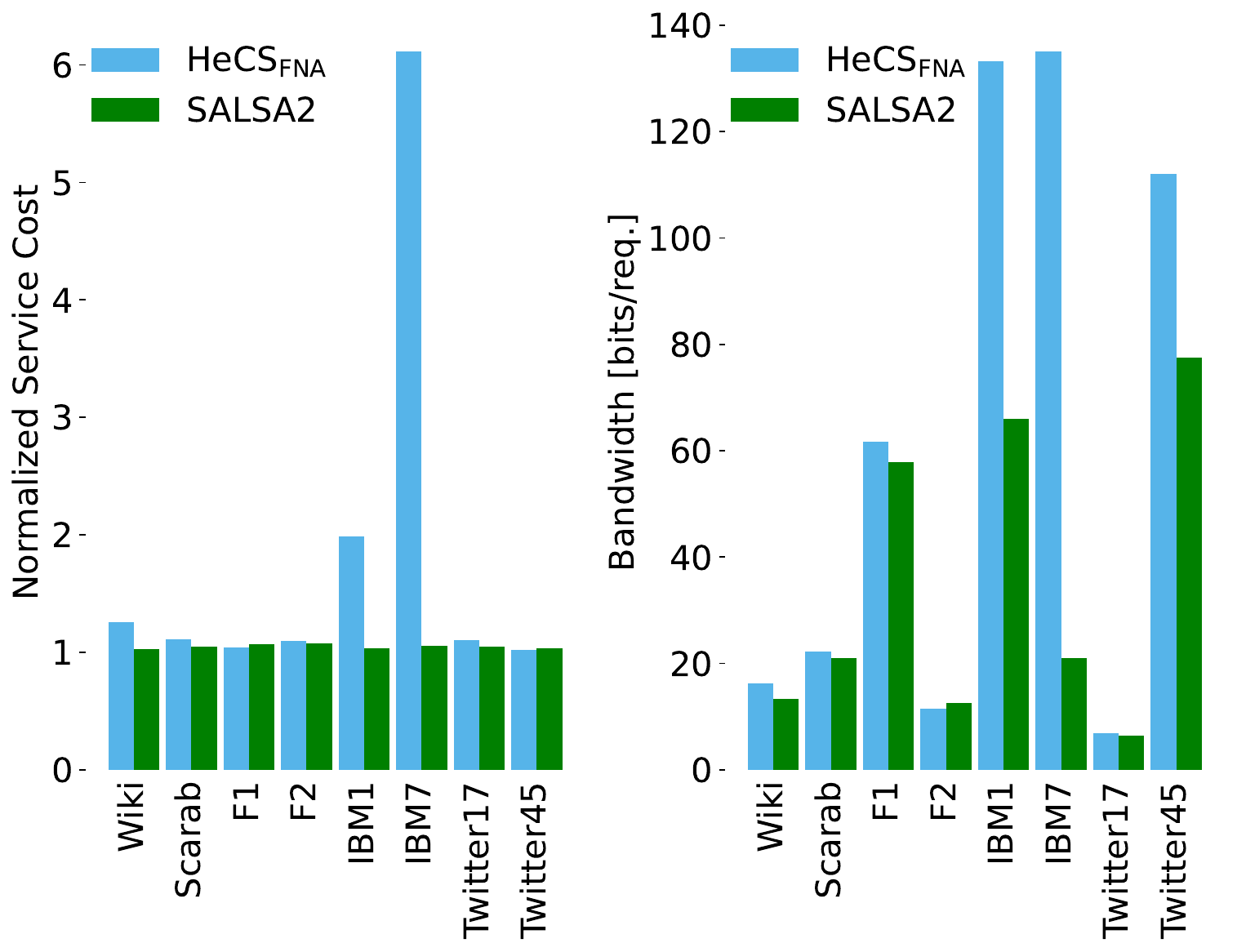}
    }
    \caption{Service cost and bandwidth consumption of the heterogeneous 3-caches baseline scenario for varying traces and miss penalty values. The costs are normalized w.r.t. hypothetical perfect indicator.}
    \label{fig:sim:hetro}
\end{figure*}

We now compare the performance of \pgmfna\ and
\salsa\ when varying the miss penalty and the cache size. We vary the miss penalty between 10, one order of magnitude higher than the cheapest cache's access cost, and 300, two orders of magnitude above the highest cache access cost. 

The results in Fig.~\ref{fig:sim:hetro} show that \salsa's service cost is either similar or significantly lower than that of \pgmfna\ across the board. \salsa\ exhibits its highest normalized costs when the miss penalty is 10 because, in this case, the respective cost of unnecessary cache access is significant (cache access cost of 1 to 3, compared to the miss penalty of 10). In most scenarios, \salsa's service cost is nearly optimal (normalized service cost close to 1). \pgmfna, on the other hand, is very sensitive to the workload's characteristics. \pgmfna's service costs is especially high for IBM1 and IBM7, which exhibit high recency (recall Tab.~\ref{tab:traces}). A possible reason is that \pgmfna's inaccurate negative exclusion probability estimation (recall Sec.~\ref{sec:epe:mrzero}), translates to multiple false-negative events.
\salsa's bandwidth is always similar or lower than that of \pgmfna, saving up to 84\% of the bandwidth (in the 64K cache, $\missp=300$, IBM7 scenario).

\subsection{Increasing the number of caches}
We now study the effect of the number of caches on \salsa's performance. We set the access cost of all caches to 2 (the average access cost of the caches in the heterogeneous scenario). The miss penalty is 30, and the capacity of each cache is 16K items.

\begin{figure*}
    \captionsetup[subfloat]{labelformat=empty}
    \centering
    \subfloat[Number of Caches = 3]{
        \includegraphics[width=\subfloatWidth]{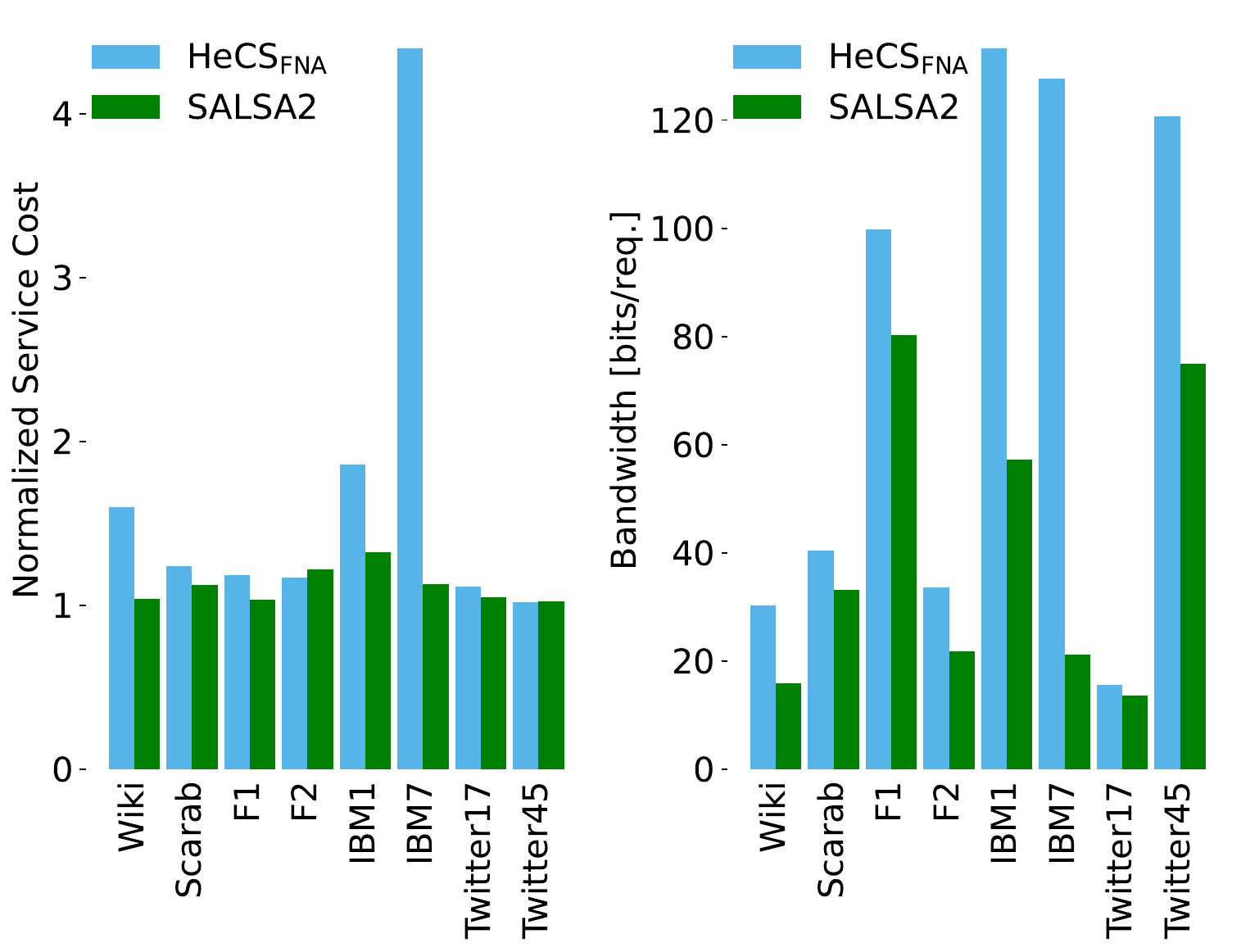}
    }
    \subfloat[Number of Caches = 6]{
        \includegraphics[width=\subfloatWidth]{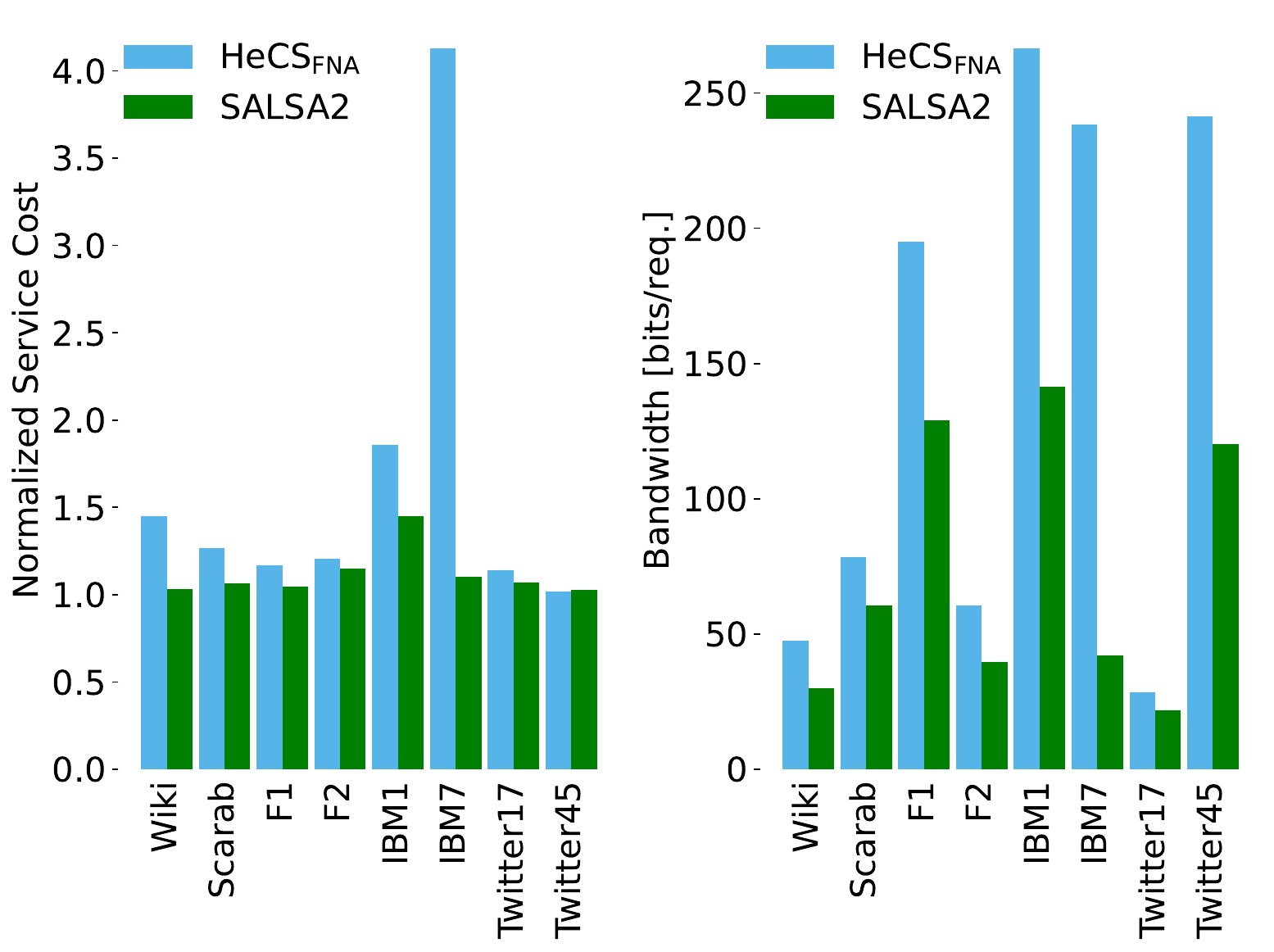}
    }
    \subfloat[Number of Caches = 9]{
        \includegraphics[width=\subfloatWidth]{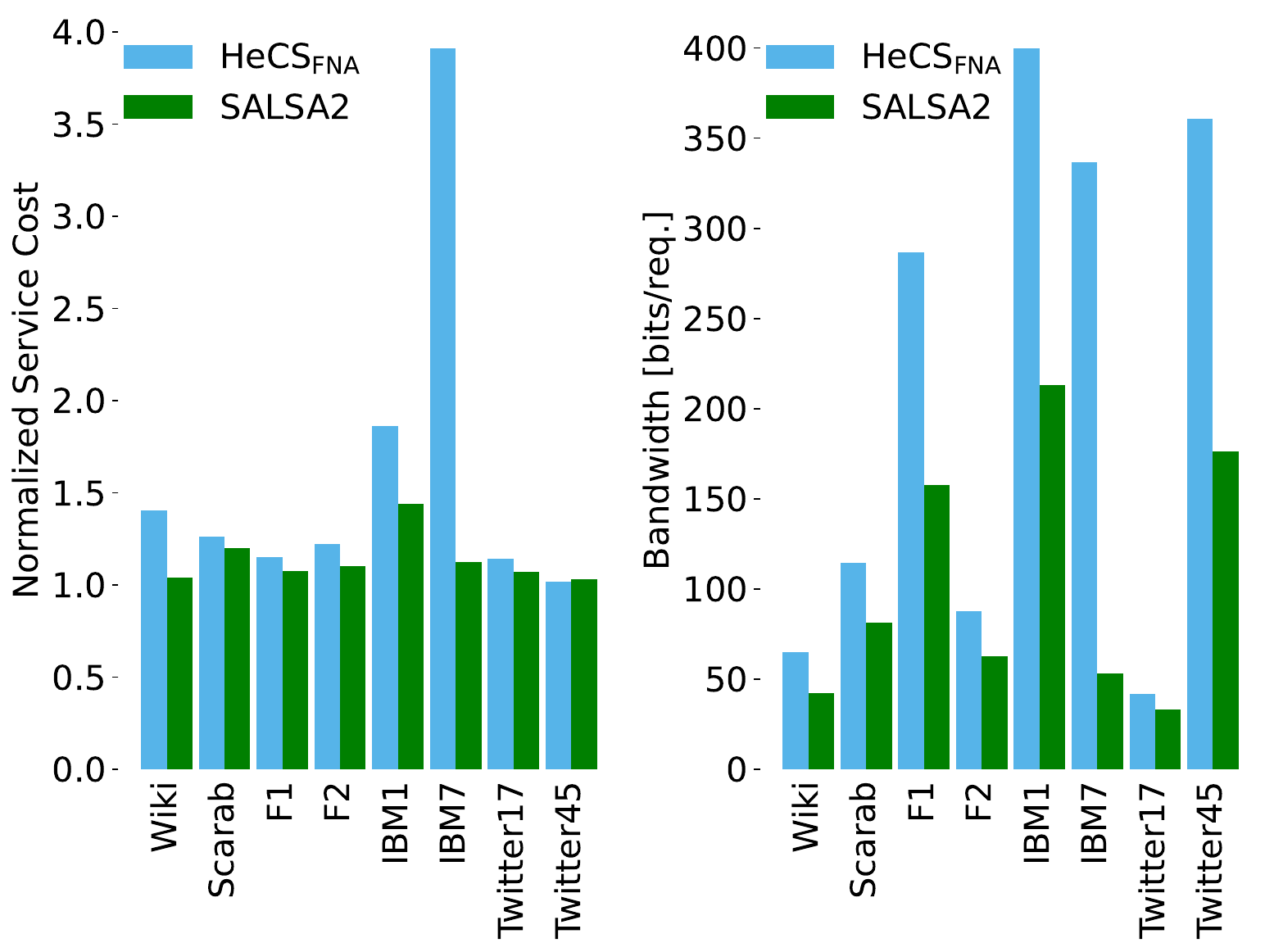}
    }
    \caption{Service cost and bandwidth consumption when increasing the number of caches. The costs are normalized w.r.t. hypothetical perfect indicator (\pif).}
    \label{fig:sim:homo}
\end{figure*}

The results in Fig.~\ref{fig:sim:homo} show in all the considered scenarios, \salsa\ obtains service cost similar or lower than that of \pgmfna, while saving up to 84\% of the bandwidth (the maximal saving is in the IBM7 9 caches scenario).

\section{Conclusion}\label{sec:conclusion}

In this paper, we introduced \salsa, a Scalable Adaptive and Learning-based Selection and Advertisement Algorithm, designed to optimize cache content advertisement and selection.
\salsa\ leverages a lightweight learning method to estimate the exclusion probabilities while considering inter-cache dependencies. Furthermore, \salsa\ dynamically scales the size and frequency of indicator advertisements, ensuring high accuracy with minimal transmission overhead, while considering the channel's reliability. 
Our extensive evaluation using a diverse set of real-world cache traces demonstrates the significant advantages of \salsa. The simulation results show that \salsa\ consistently outperforms existing solutions in terms of service cost and bandwidth consumption. 
In particular, \salsa's normalized service cost is nearly optimal across most scenarios, and the bandwidth consumption is reduced by up to 84\% 
compared to the state-of-the-art solution. These results underscore \salsa's capability to substantially improve cache advertisement and selection, making it a robust and versatile solution for modern distributed caching systems challenges.
\bibliographystyle{IEEEtran}
\bibliography{Refs}

\appendix\label{app}
In this appendix, we detail the procedures xmtDeltaIsCheaper() and estimatedBW(), which appear in Alg.~\ref{alg:after_insertion:full} and Alg.~\ref{alg:after_insertion:delta}, respectively.

\paragraph*{xmtDeltaIsCheaper()} Let $\indSize$ denote the size of the indicator when the function is called. Keeping in full indicator mode during the next synchronization period translates to transmitting $R \cdot \indSize$ bit in total. To consider delta mode, the system retains both an updated indicator and a stale indicator that reflects the last-advertised indicator~\cite{FN_aware_ToN, CAB}. 
Denote by $D$ the {\em diff}, namely, the number of bits toggled in the indicator since the last advertisement. Explicitizing the address of a single flipped bit in the indicator requires $\ceil{\log \indSize}$ bits. Hence, the expected bit-count of a single delta update is $D \cdot \ceil{\log \indSize}$ bits. 
Assuming that the average pace of bits flipping in the indicator will not change shortly, the estimated overall number of bits to be transmitted during the next synchronization period is $R \cdot D \cdot \ceil{\log \indSize}$ for the delta updates, plus a single full-indicator $\indSize$-bits synchronization update at the end of the period. It follows that xmtDeltaIsCHeaper() returns True if and only if $\period \cdot D \cdot \ceil{\log \indSize} + \indSize < \period \cdot \indSize$.

\paragraph*{estimatedBw} Let $D$ denote the average number of flipped bits reported by a delta update during the last synchronization period. Let $\indSize$ denote the current counter size and $\indSize^j$ denote an optional alternative indicator size. Then, scaling the indicator from $\indSize$ to $\indSize^j$ is expected to scale the diff size to $\frac{\indSize_j}{\indSize} D$. The number of bits required to represent the address of each flipped bit is $\log \indSize_j$. It follows that for an 
$\indSize_j$-bits indicator, the average number of bits required to be transmitted at each delta update is $\frac{\indSize^j}{\indSize} D \cdot \ceil {\log \indSize^j}$. The bandwidth budget required to transmit such update once in $\uIntervalMinFeasible$ insertions is $\frac{\indSize_j}{\indSize \cdot \uIntervalMinFeasible} D \cdot \ceil {\log \indSize^j}$. In addition, the cache will advertise a $\indSize^j$-bits full indicator once in $\period \cdot \uInterval$ insertions. Combining the reasonings above, the estimated bandwidth budget of operating an $\indSize^j$-bits indicator's delta update is 
$$
\textrm{estimatedBw}(\indSize^j) = 
\frac{\indSize^j}{\indSize \cdot \uIntervalMinFeasible} D \cdot \ceil{\log \indSize_j} + \frac{\indSize^j}{\period \cdot \uInterval}.
$$

\end{document}